\documentclass[twocolumn,10pt,a4paper,tightenlines,lengthcheck]{iopart}
\usepackage{graphicx}
\usepackage[breaklinks=true,colorlinks=true,linkcolor=blue,urlcolor=blue,citecolor=blue]{hyperref}
\usepackage{times}
\usepackage{subfigure}
\usepackage{epsfig}
\usepackage{epstopdf}
\begin{document}
\title[Sharply tunable group velocity in alkali vapors]{Sharply tunable group velocity in alkali vapors using a single low-power control field}
\author{Pardeep Kumar$^{\dagger}$ and Shubhrangshu Dasgupta}
\date{\today}
\address{{Department of Physics,
Indian Institute of Technology Ropar, Nangal Road, Rupnagar, Punjab, India $140001$}}
\ead{pradeep.kumar@iitrpr.ac.in}

\begin{abstract}
We show how a single linearly polarized control field can produce a sharply tunable group velocity of a weak probe field at resonance in a four-level atomic configuration of alkali vapors. The dispersion can be switched from normal to anomalous along with vanishing absorption, just by changing intensity of the resonant control field. In addition, by allowing different intensities of the different polarization components of the control field, the anomalous dispersion can be \textit{switched back to the normal.} This thereby creates a ``valley of anomaly" in group index variation and offers two sets of control field intensities, for which the system behaves like a vacuum. The explicit analytical expressions for the probe coherence are provided along with all physical explanations. We demonstrate our results in $J=\frac{1}{2}\leftrightarrow J=\frac{1}{2}$ transition for $D_1$ lines in alkali atoms, in which one can obtain a group index as large as 3.2$\times 10^{8}$  and as negative as -1.5$\times 10^{5}$ using a control field with power as low as 0.017 mW/cm$^{2}$ and 9.56 mW/cm$^2$, respectively.
\end{abstract}

\section{Introduction}
The study of subluminal and superluminal propagation of light has been a focus of research in last few decades. A considerable work has been done to control slow and fast light by making use of quantum interference in multilevel atomic system coupled with several fields. Harris \textit{et. al.} \cite{harris1992} suggested that electromagnetically induced transparency (EIT) \cite{harris1997} can be used to obtain group velocity much smaller than $c$, the speed of light in vacuum. A number of impressive experiments \cite{harris1999,chien2001,wang2000,kasapi1995,schmidt1996,kash1999,budker1999} has been performed on the group velocity of optical pulses. Group velocities as low as 17 ms$^{-1}$ in a cloud of sodium atoms below the transition temperature for Bose-Einstein condensation \cite{harris1999} and those of the order of 90 ms$^{-1}$ in optically dense hot rubidium gas \cite{kash1999} have been reported.  The laser pulses could also be stopped \cite{chien2001}. On the other hand, superluminal propagation of optical pulses has been demonstrated  by Wang \textit{et. al.} \cite{wang2000}. Sommerfeld and Brillouin \cite{brillouin1960} theoretically showed that anomalous nature of dispersion profile may result in a group velocity faster than $c$, thanks to the wave nature of light \cite{born1997}. An atomic medium can further be optically manipulated in a controlled fashion to switch from subluminal group velocities to the superluminal. Such a control using a microwave field between two lower metastable states in a $\Lambda$-configuration was proposed by Agarwal and coworkers \cite{aggarwal2001}. Similar switching can be done by adjusting the relative phase of the weak probe fields applied to a V-type three-level system \cite{arbiv2001} and the phase-difference between the strong control fields \cite{sahrai2004}. Tunability of group velocity in a V-system with indirect incoherent pumping has been further discussed in \cite{saaidi2011}.

 In addition, the studies of coherence in systems with more than three energy levels \cite{Li2007,Zhang2007,Yan2001,harris1998}  have attracted much interest. Among these, four-level systems with N-configuration may be specially mentioned, in which the power requirement of the control laser field is quite less, compared to that in the other configurations, to control group velocities of a probe field. It was proposed by Harris in his seminal paper \cite{harris1998} and later demonstrated in \cite{Kang2004}  that a second control field in an EIT system may lead to {\it absorption of the probe field} at resonance (contrary to transparency, as in usual EIT phenomenon), while it switches the dispersion from normal to anomalous. Optical properties of Rubidium in a Doppler-broadened N-type configuration of $D_2$ transition have been discussed by Chen \textit{et. al.} \cite{Chen2009}. Salloum \textit{et. al.} \cite{Salloum2010,Salloum2011} have explained the variations of dispersion and transparency of probe field in N-configuration with the help of various resonance terms.

In this paper, we consider a four-level configuration [Figure 1] as in $J=\frac{1}{2}\leftrightarrow J=\frac{1}{2}$ transition for $D_1$ lines in alkali atoms \cite{s2006}, which can be understood as a combination of two N-configurations, each with a single circularly polarized control field in action. We show the possibility of switching the propagation from subluminal to superluminal domain, along with zero absorption, by tuning the strength of a {\it single} resonant linearly polarized control field. A very large group index($\sim$ 3.2$\times 10^{8}$) can be obtained which results in ultraslow light. Note that in \cite{sankar2013}, a similar switching has been reported in a different four-level configuration that requires {\it two} control fields, unlike in the present case. We emphasize that in the present system, the control field does not lead to probe absorption, as opposed to that reported in the previous studies of N-systems. Therefore, this configuration can be considered as a preferable candidate to obtain sharp tunability of group velocity.  The polarization components of control field in two transitions dress the system. For a particular set of these components of control  field, the transition from the excited dressed state manifold to the ground state manifold create a gain doublet, which results in superluminality. For a different set of control field components, the absorption from the ground state doublet dominates over the gain, and results in subluminality.  We further emphasize that the power of the control field can be kept much lower (of the order of $10^{-3}$mW/cm$^{2}$) than those reported in the early experiments on group velocity manipulation, while maintaining the tunability of group velocities. This is the first time, the possibility of changing dispersion from normal to anomalous and \textit{then back to normal} and thereby of producing a valley of anomalous dispersion using such a low power control field is proposed.

The organization of the paper is as follows. In Section 2, we describe the atomic configuration along with relevant density matrix equations. We also propose a possible experimental set-up  to achieve such configuration. In Section 3, we provide the exact analytical solution of these equations, up to first order of the probe field. We here describe how the group index can be made tunable with respect to the control field strength. We discuss the effect of Doppler broadening in Section 4. In Section 5,  the physical scenario behind our results is described. The Section 6 highlights the conclusion.

\section{Model}

\begin{figure}[ht!]
\centering
\includegraphics[scale=0.35]{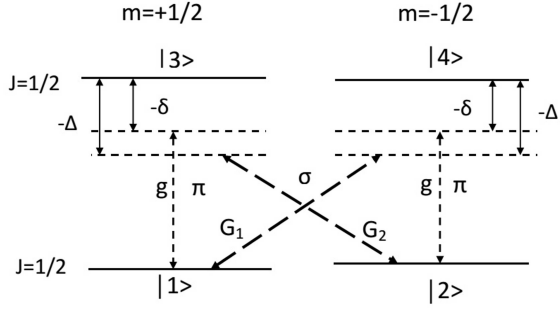}
\label{Fig.1}
\caption{Schematic energy-level structure of a  four-level atomic system involving $J=\frac{1}{2}\leftrightarrow J=\frac{1}{2}$ transition. The $\sigma$-polarized control field components with Rabi frequencies $2G_1$ and $2G_2$ induce the transitions $\vert1\rangle\leftrightarrow\vert4\rangle$ and $\vert2\rangle\leftrightarrow\vert3\rangle$ while a $\pi$-polarized probe field with Rabi frequency $2g$ interacts with the transitions $\vert1\rangle\leftrightarrow\vert3\rangle$ and $\vert 2\rangle\leftrightarrow\vert 4\rangle$. The parameters $\Delta$ and $\delta$ are the control and probe field detunings, respectively.}

\end{figure}

\begin{figure}[ht!]
\centering
\includegraphics[scale=0.4]{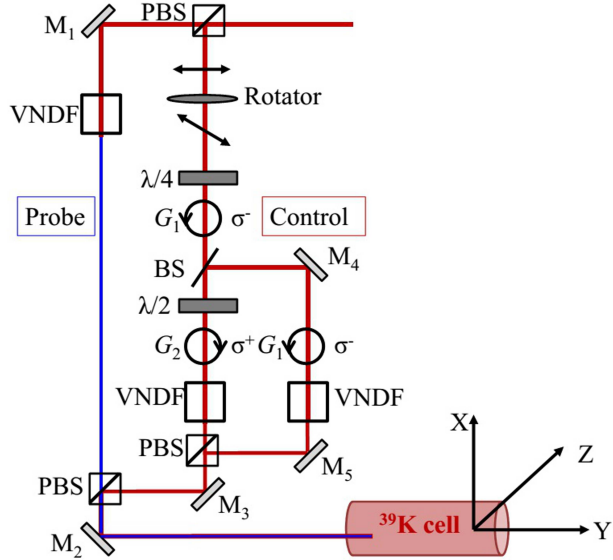}
\caption{Schematic experimental setup. PBS, polarizing beam splitter; VNDF, variable neutral density filter; $\lambda$/4, quarter-wave plate; $\lambda$/2, half-wave plate; BS, 50:50 beam splitter; M$_{i}$'s are 100\% reflecting mirrors. The VNDFs can be used to control the intensities of the circular components of the control fields.}
\label{fig(5)}
\end{figure}

We choose a generic four-level configuration, as shown in Figure 1, that comprises of two ground states $|1\rangle$, $|2\rangle$ and two excited states $|3\rangle$, $|4\rangle$. The D$_1$ transition (4S$_{1/2}$ $\leftrightarrow$ 4P$_{1/2}$) in ${}^{39}$K atoms may be identified with such energy level configuration. The spin of ${}^{39}$K  nucleus can be ignored for theoretical simplifications \cite{brown1997}, so that 4S$_{1/2}$ and 4P$_{1/2}$ states consist of two degenerate Zeeman sublevels with magnetic quantum numbers $m=\pm1/2$. The transitions $\vert1\rangle\leftrightarrow\vert3\rangle$ and $\vert2\rangle\leftrightarrow\vert4\rangle$ are driven simultaneously by a weak linearly polarized electromagnetic field $\vec{E_p}=\hat{z}\varepsilon_pe^{-i(\omega_pt-ky)}+$c.c.. The Rabi frequency for this probe field is given by $2g=2\left(\frac{\vec{d_{31}}.\hat{z}\varepsilon_p}{\hbar}\right)=2\left(\frac{\vec{d_{42}}.\hat{z}\varepsilon_p}{\hbar}\right)$, where $\vec{d}_{ij}$ is the electric dipole moment matrix element between the levels $\vert i\rangle$ and $\vert j\rangle$. A strong linearly polarized control field  $\vec{E_c}=\hat{x}\varepsilon_ce^{-i(\omega_ct-ky)}+$c.c. is applied to drive the transitions $\vert1\rangle\leftrightarrow\vert4\rangle$ and $\vert2\rangle\leftrightarrow\vert3\rangle$. The corresponding Rabi frequencies are given by $2G_1=2\left(\frac{\vec{d_{41}}.\hat{x}\varepsilon_c}{\hbar}\right)$ and $2G_2=2\left(\frac{\vec{d_{32}}.\hat{x}\varepsilon_c}{\hbar}\right)$, respectively. It should be borne in mind that in the previous studies on N-systems \cite{harris1998,Kang2004,Chen2009,Salloum2010,Salloum2011}, only one atomic transition is driven by the probe field, while both the control fields share a common ground state. This is unlike the system in this paper, in which the control field do not share any of the atomic states. A possible experimental setup for this configuration \cite{santosh2013} is shown in Figure 2. 

Note that the configuration of Figure 1 could also be understood as a combination of two $\Lambda$-systems, each of which is composed of one control field and one of the probe field, with degenerate excited state and ground state manifold. So, the total coherences of the $\pi$-polarized field is a sum of the coherences, as calculated for each of these subsystems. Such system has been used to obtain {\it nonlinear} parametric amplification of the probe fields \cite{lukin} and {\it nonlinear} two-color lasing \cite{wu}, while our model generates {\it linear} probe gain, as discussed later in Section 4.

The Hamiltonian for this system in the dipole approximation can be written as
\begin{eqnarray}
\begin{array}{ccc}
\hat{H}&=& \hbar \left[\omega_{21}\vert2\rangle\langle2\vert+\omega_{31}\vert3\rangle\langle3\vert+\omega_{41}\vert4\rangle\langle4\vert \right]\\
& &  -\left[(\vec{d_{41}}\vert4\rangle\langle1\vert+\vec{d_{32}}\vert3\rangle\langle2\vert+h.c.).\vec{E_c}\right]\\
& &-\left[(\vec{d_{31}}\vert3\rangle\langle1\vert+\vec{d_{42}}\vert4\rangle\langle2\vert+h.c.).\vec{E_p}\right]\;.
\end{array}
\label{eq1}
\end{eqnarray}
Here zero of energy is defined at the level $\vert1\rangle$ and $\hbar\omega_{\alpha\beta}$ is the energy difference between the levels $\vert \alpha\rangle$ and $\vert \beta\rangle$.
Using the Markovian master equation with natural decay terms, the following density matrix equations are obtained:
\begin{eqnarray}
\begin{array}{lll}
\dot{\tilde{\rho}}_{11} &=& \gamma_{13}\tilde{\rho}_{33}+\gamma_{14}\tilde{\rho}_{44}+i(G_1^\ast\tilde{\rho}_{41}-G_1\tilde{\rho}_{14})+i(g^{\ast}\tilde{\rho}_{31}e^{i\omega_{pc}t}-g\tilde{\rho}_{13}e^{-i\omega_{pc}t})\\
\dot{\tilde{\rho}}_{33} &=& -(\gamma_{13}+\gamma_{23})\tilde{\rho}_{33}+i(G_2\tilde{\rho}_{23}-G_2^\ast\tilde{\rho}_{32})+i(g\tilde{\rho}_{13}e^{-i\omega_{pc}t}-g^\ast\tilde{\rho}_{31}e^{i\omega_{pc}t})\\
\dot{\tilde{\rho}}_{44} &=& -(\gamma_{14}+\gamma_{24})\tilde{\rho}_{44}+i(G_1\tilde{\rho}_{14}-G_1^\ast\tilde{\rho}_{41})+i(g\tilde{\rho}_{24}e^{-i\omega_{pc}t}-g^\ast\tilde{\rho}_{42}e^{i\omega_{pc}t})\\
\dot{\tilde{\rho}}_{31} &=& -i(\Delta -i \Gamma_{31})\tilde{\rho}_{33}+i(G_2\tilde{\rho}_{21}-G_1\tilde{\rho}_{34})+i(\tilde{\rho}_{11}-\tilde{\rho}_{33})ge^{-i\omega_{pc}t}\\
\dot{\tilde{\rho}}_{32} &=& -i(\Delta -i \Gamma_{32})\tilde{\rho}_{32}+i(1-\tilde{\rho}_{11}-2\tilde{\rho}_{33}-\tilde{\rho}_{44})G_2+i(\tilde{\rho}_{12}-\tilde{\rho}_{34})ge^{-i\omega_{pc}t}\\
\dot{\tilde{\rho}}_{43} &=& - \Gamma_{43}\tilde{\rho}_{43}+i(G_1\tilde{\rho}_{13}-G^\ast _2\tilde{\rho}_{42})+i(g\tilde{\rho}_{23}e^{-i\omega_{pc}t}-g^\ast \tilde{\rho}_{41}e^{i\omega_{pc}t})\\
\dot{\tilde{\rho}}_{42} &=& -i(\Delta -i \Gamma_{42})\tilde{\rho}_{42}+i(G_1\tilde{\rho}_{12}-G_2\tilde{\rho}_{43})+i(1-\tilde{\rho}_{11}-\tilde{\rho}_{33}-2\tilde{\rho}_{44})ge^{-i\omega_{pc}t}\\
\dot{\tilde{\rho}}_{41} &=& -( i\Delta +\Gamma_{41})\tilde{\rho}_{41}+i(\tilde{\rho}_{11}- \tilde{\rho}_{44})G_1+i(\tilde{\rho}_{21}- \tilde{\rho}_{43})ge^{-i\omega_{pc}t}\\
\dot{\tilde{\rho}}_{21} &=& - \Gamma_{21}\tilde{\rho}_{21}+i(G_2^\ast \tilde{\rho}_{31}-G_1 \tilde{\rho}_{24})+i(g^\ast\tilde{\rho}_{41}e^{i\omega_{pc}t}-g\tilde{\rho}_{23}e^{-i\omega_{pc}t})\;,
\end{array}
\label{eq2}
\end{eqnarray}

where $\Delta=\omega_{c}-\omega_{41}$ ($\Delta=\omega_{c}-\omega_{32}$)  is the detuning of the control field from the transition $\vert 1\rangle \leftrightarrow \vert 4 \rangle$ ($\vert 2\rangle \leftrightarrow \vert 3 \rangle$) , $\delta=\omega_{p}-\omega_{42}$ ($\delta=\omega_{p}-\omega_{31}$) is the detuning of the probe field from the transition $\vert 2 \rangle \leftrightarrow \vert4 \rangle$ ($\vert 1 \rangle \leftrightarrow \vert 3 \rangle$), $\omega_{pc}=\omega_p-\omega_c=\delta-\Delta$ is the frequency difference of the probe and control fields, the spontaneous emission rate from the level $\vert j\rangle$ to $\vert i\rangle$ is $\gamma_{ij}$, the de-phasing rate of coherence between the levels $\vert j\rangle$ and $\vert i\rangle$ is $\Gamma_{ij}=\frac{1}{2}\sum_{k}(\gamma_{ki}+\gamma_{kj})+\gamma_{\rm coll}$, where $\gamma_{coll}$ is the collisional decay rate. Here onwards, we choose $\gamma_{24}=\gamma_{13}=\gamma_{1}$, $\gamma_{14}=\gamma_{23}=\gamma_{2}$,
 $\Gamma_{32}=\Gamma_{41}=\Gamma_{31}=\Gamma_{42}=\Gamma_{1}=\frac{1}{2}(\gamma_{1}+\gamma_{2})$, $\Gamma_{21}=\Gamma_{2}=0$, $\Gamma_{43}=\Gamma_{3}=(\gamma_{1}+\gamma_{2})$ and $\gamma_{\rm coll}=0$.
The above equations are obtained by using rotating wave approximation so as to neglect the highly oscillating terms. The transformations for the density matrix elements are as follows: $\rho_{31} = \tilde{\rho}_{31}e^{-i\omega_{c}t}$,    $\rho_{32} = \tilde{\rho}_{32}e^{-i\omega_{c}t}$, $\rho_{42} = \tilde{\rho}_{42}e^{-i\omega_{c}t}$,  $\rho_{41} = \tilde{\rho}_{41}e^{-i\omega_{c}t}$. The rest of the elements remain the same.

To solve  (\ref{eq2}) in steady state, we expand the density matrix elements in Fourier domain in terms of $\omega_{pc}$ as

\begin{equation}
\tilde{\rho}_{\alpha\beta}=\tilde{\rho}_{\alpha\beta}^{(0)}+ge^{-i\omega_{pc}t}\tilde{\rho}_{\alpha\beta}^{(-1)}+g^\ast e^{i\omega_{pc}t}\tilde{\rho}_{\alpha\beta}^{(+1)} \;.
\label{eq3}
\end{equation}

Comparing the coefficients  of $ e^{\pm i \omega_{pc}t} $, we obtain a set of  linear algebraic equations for $\tilde{\rho}_{\alpha \beta}^{(n)}$ ($n\in \pm 1, 0$). We solve these equations for different values of $n$, as described in the next section.

The coherence for the $\pi$-polarized probe field is determined by two contributions \cite{hashmi2008}, one from $\tilde{\rho}_{31}^{(-1)}$ and other from $\tilde{\rho}_{42}^{(-1)}$ and is given by:
\begin{equation}
\rho_{\pi}=\tilde{\rho}_{31}^{(-1)}+\tilde{\rho}_{42}^{(-1)}\;.
\label{eq4}
\end{equation}
The response of the atomic system to the applied fields is determined by the susceptibility $\chi_{\pi}$, which is defined as
\begin{equation}
\chi_{\pi}=\frac{3Nc^3}{2\omega_{p}^3}\rho_{\pi}\;,
\label{eq5}
\end{equation}
where $N$ is the number density of the atomic medium. The real and imaginary parts of $\chi_{\pi}$ correspond to dispersion and absorption of the probe field, respectively.
\section{Results}

\subsection{Case when $G_1=G_2=G$}

In the following, we will first consider the pump Rabi frequencies to be equal, i.e., $G_1=G_2=G$. This is justified because, the dipole moments in the transitions $|1\rangle\leftrightarrow |4\rangle$ and $|2\rangle\leftrightarrow |3\rangle$ are the same and the same linearly polarized control field interacts with these transitions.


\indent
When the probe field is switched off, the system behaves like two two-level systems. The atomic population (zeroth order in terms of the probe fields) gets redistributed among all the energy levels and are given as:

\begin{eqnarray}
\begin{array}{lll}
\tilde{\rho}_{11}^{(0)}&=\tilde{\rho}_{22}^{(0)}&=\frac{\gamma_1+\gamma_2+x}{2(\gamma_{1}+\gamma_{2}+2x)}\;,\\
\tilde{\rho}_{33}^{(0)}&=\tilde{\rho}_{44}^{(0)}&=\frac{x}{2(\gamma_{1}+\gamma_{2}+2x)}\;,\\
\end{array}
\label{eq6}
\end{eqnarray}
where, $x = \frac{2\vert G\vert^2\Gamma_{1}}{\Delta^2+\Gamma_{1}^2}$.
The zeroth order coherence between the energy levels coupled via the control field are given by
\begin{equation}
\tilde{\rho}_{14}^{(0)}=\tilde{\rho}_{23}^{(0)}=-\frac{iG^{\ast}}{-i\Delta+\Gamma_1}\tilde{\rho}_{pd}^{(0)}\;,
\label{eq7}
\end{equation}
where,
\begin{equation}
\begin{array}{llll}
\tilde{\rho}_{pd}^{(0)}&=\tilde{\rho}_{11}^{(0)}-\tilde{\rho}_{33}^{(0)}&=\tilde{\rho}_{22}^{(0)}-\tilde{\rho}_{44}^{(0)}&=\frac{\gamma_1+\gamma_2}{2(\gamma_1+\gamma_2+2x)}\;.
\end{array}
\label{eq8}
\end{equation}
The above expressions show the two level features in the system.
\begin{figure}[!ht]
\begin{center}
\begin{tabular}{cc}
\subfigure[]{\includegraphics[scale=0.23]{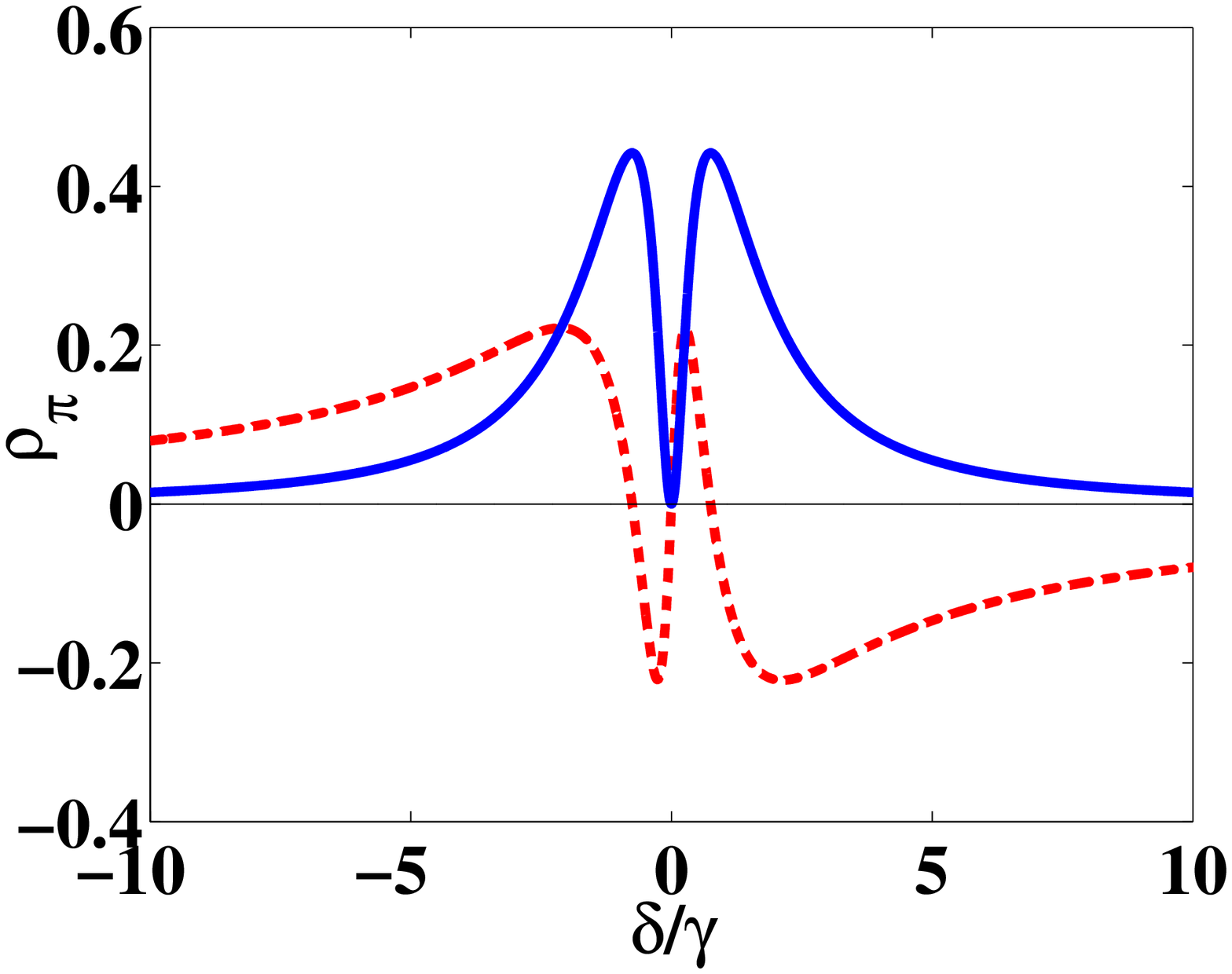}} & \subfigure[]{\includegraphics[scale=0.23]{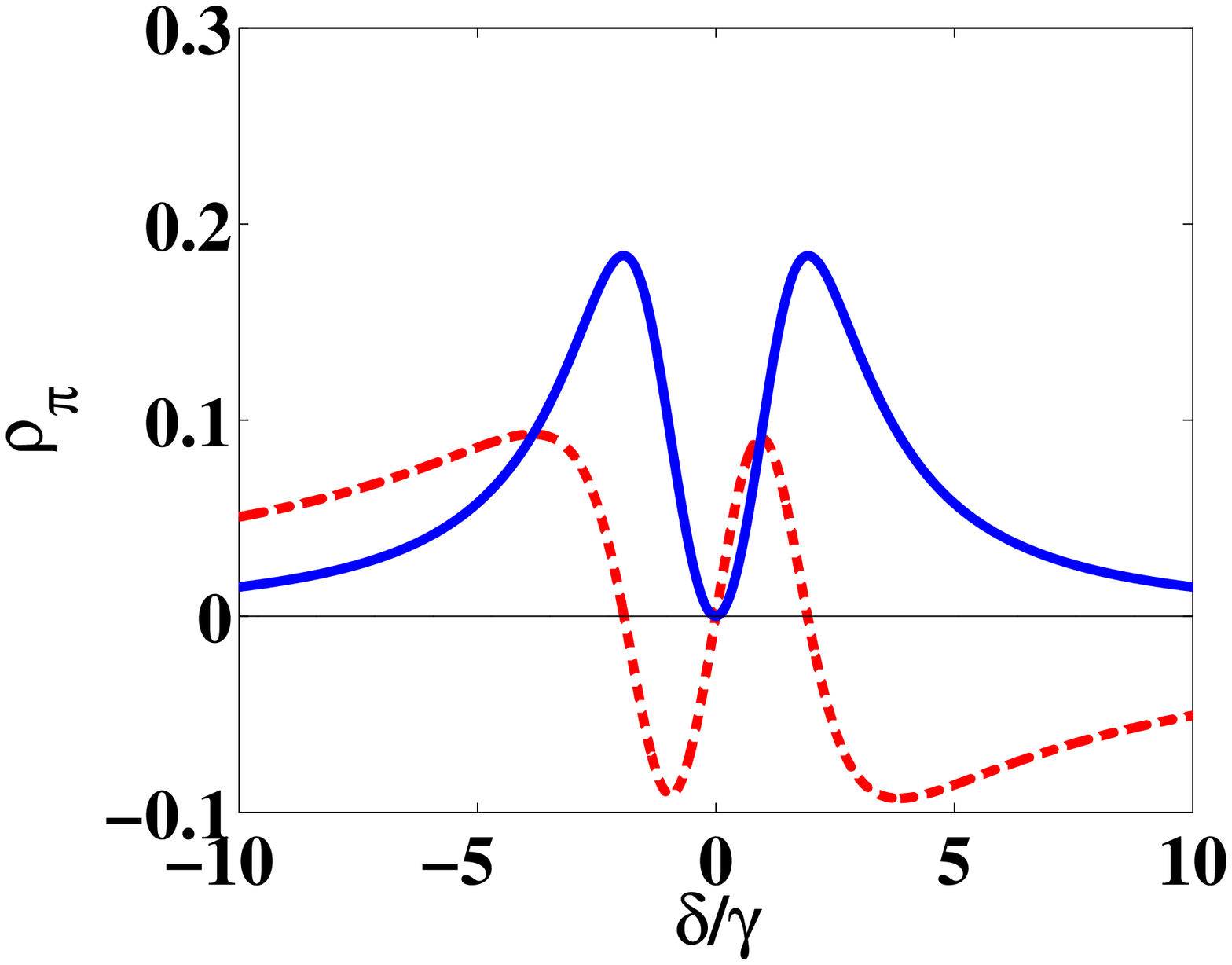}} \\ \subfigure[]{\includegraphics[scale=0.23]{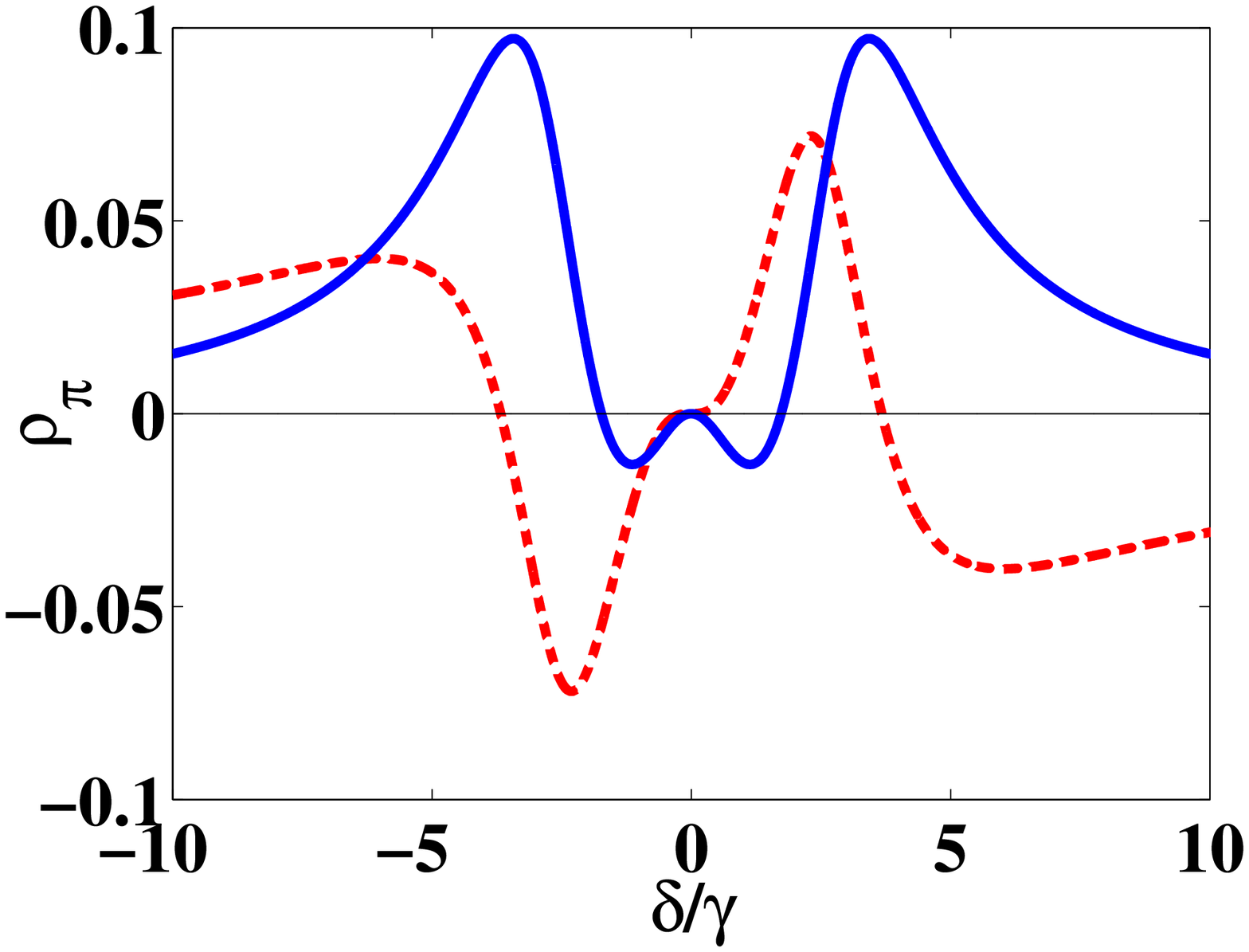}} & \subfigure[]{\includegraphics[scale=0.23]{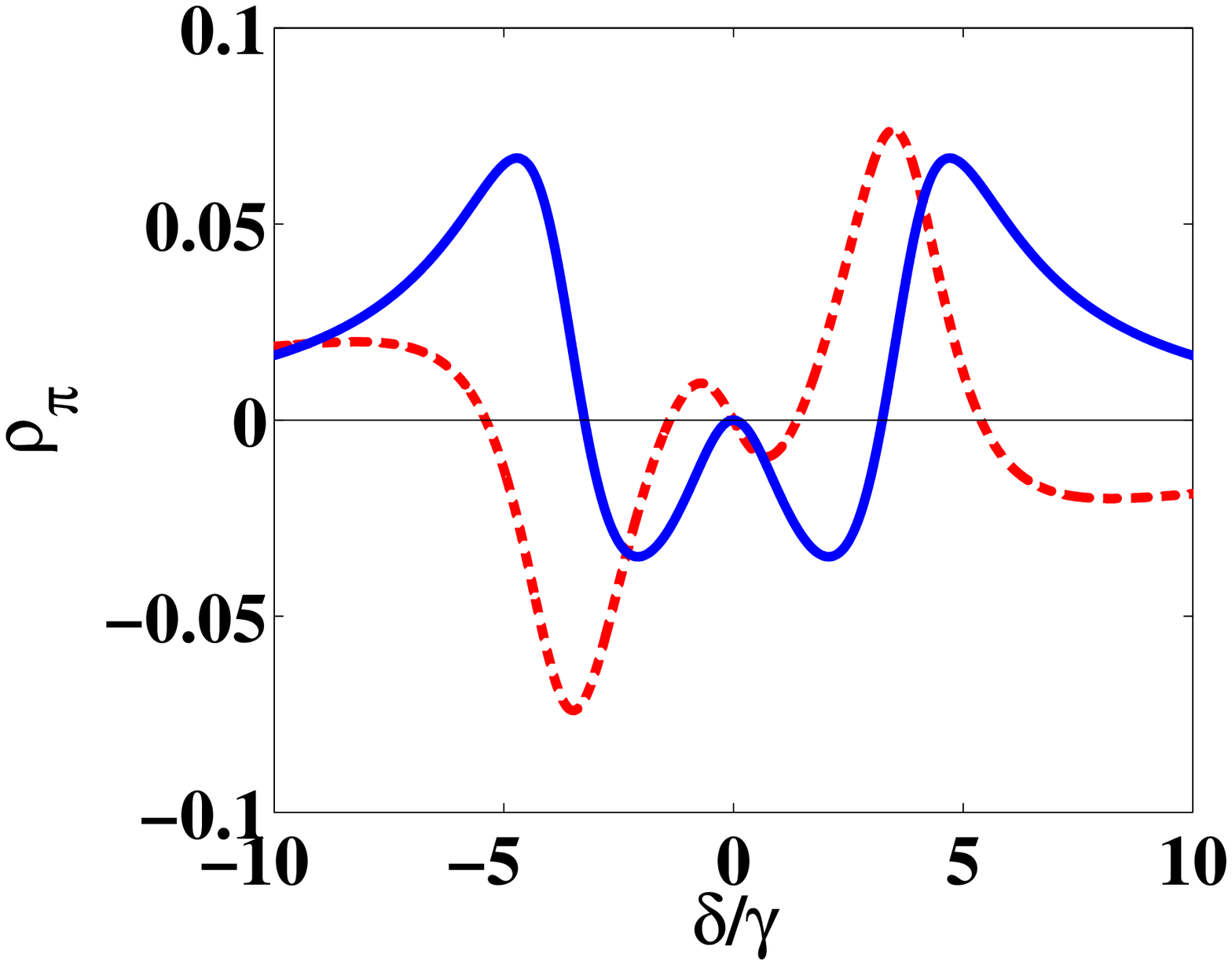}}\\
\end{tabular}
\begin{center}

\caption{Variation of real (dotted line) and imaginary parts (solid line) of $\rho_{\pi}$ with probe detuning $\delta/\gamma$ for parameters $\gamma_{13}=\gamma_{24}=2\gamma$, $\gamma_{23}=\gamma_{14}=\gamma$ and $\gamma_{\rm coll}=0$ with $\Delta=0$ and (a) $G=0.5\gamma$, (b) $\gamma$, (c) 1.5$\gamma$, (d) 2$\gamma$.}
\end{center}
\label{fig2}
\end{center}
\end{figure}

\begin{figure}[ht!]
\centering
\includegraphics[scale=0.4]{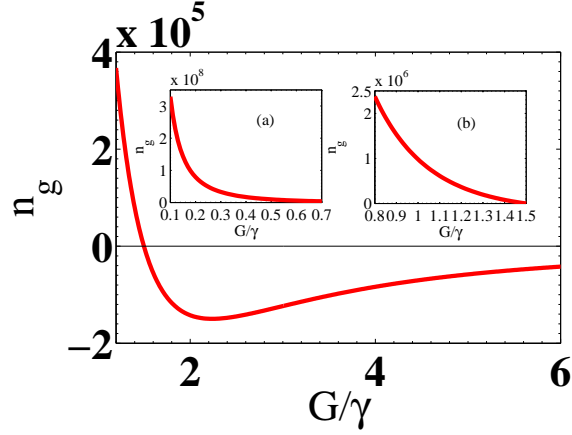}
\caption{Variation of group index $n_{g}$ with $G/\gamma$ for $N=10^{12}$ cm$^{-3}$, $\Delta=0$ and $\delta=0$. All other parameters are same as in Figure 3.}
\label{fig(3)}
\end{figure}

 \begin{figure}[!ht]
\begin{center}
\begin{tabular}{cc}
\subfigure[]{\includegraphics[scale=0.23]{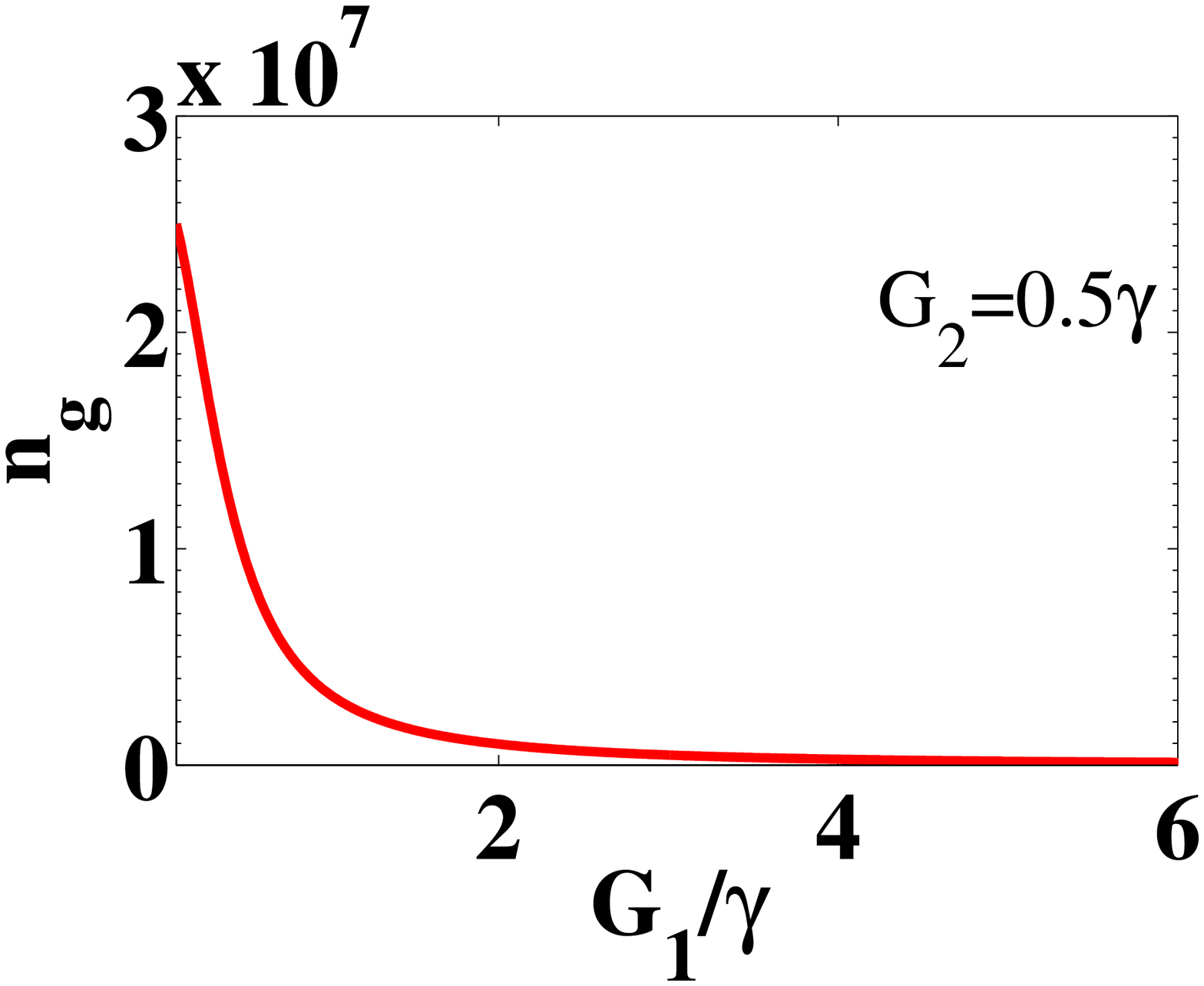}} & \subfigure[]{\includegraphics[scale=0.23]{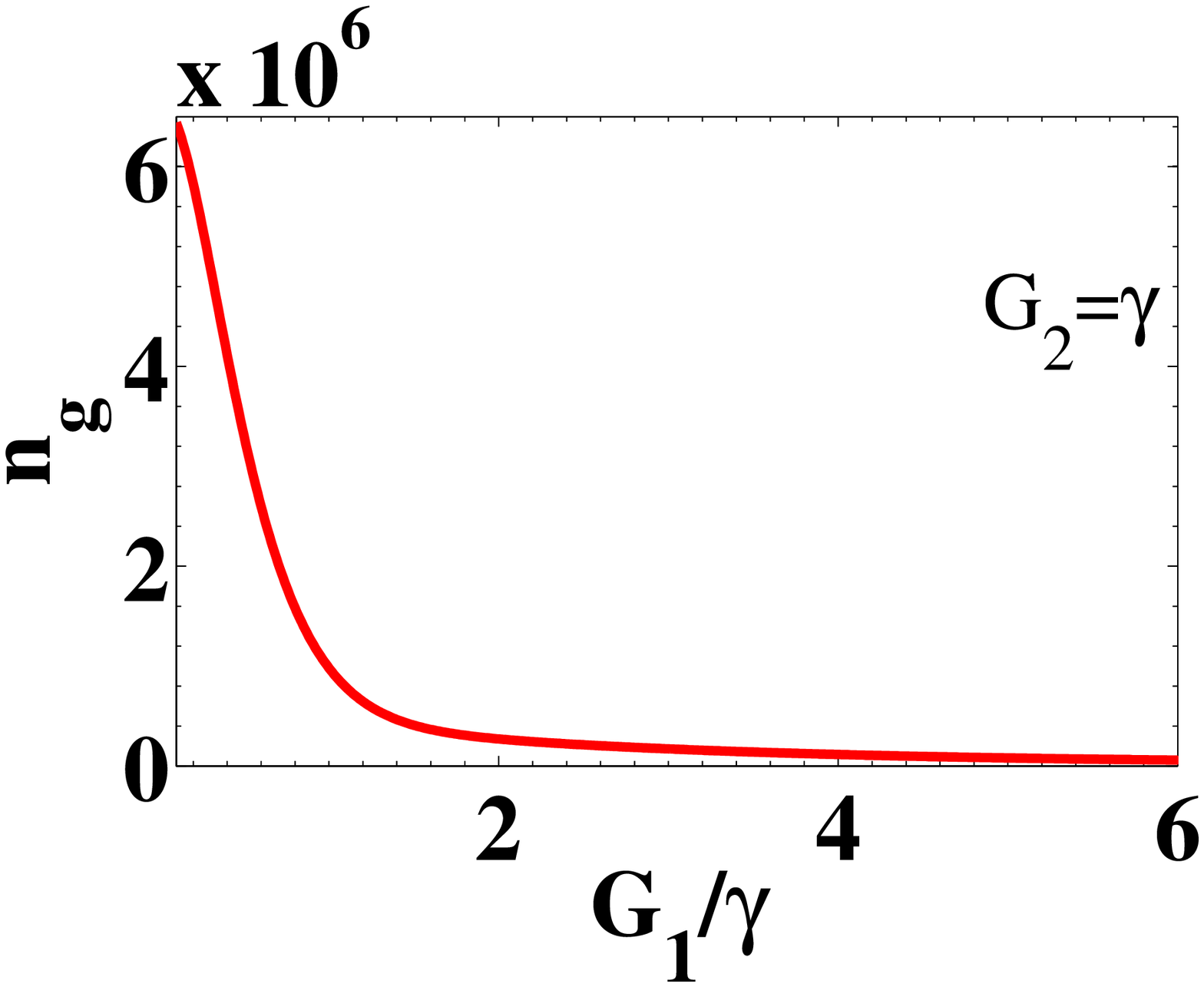}}\\ \subfigure[]{\includegraphics[scale=0.23]{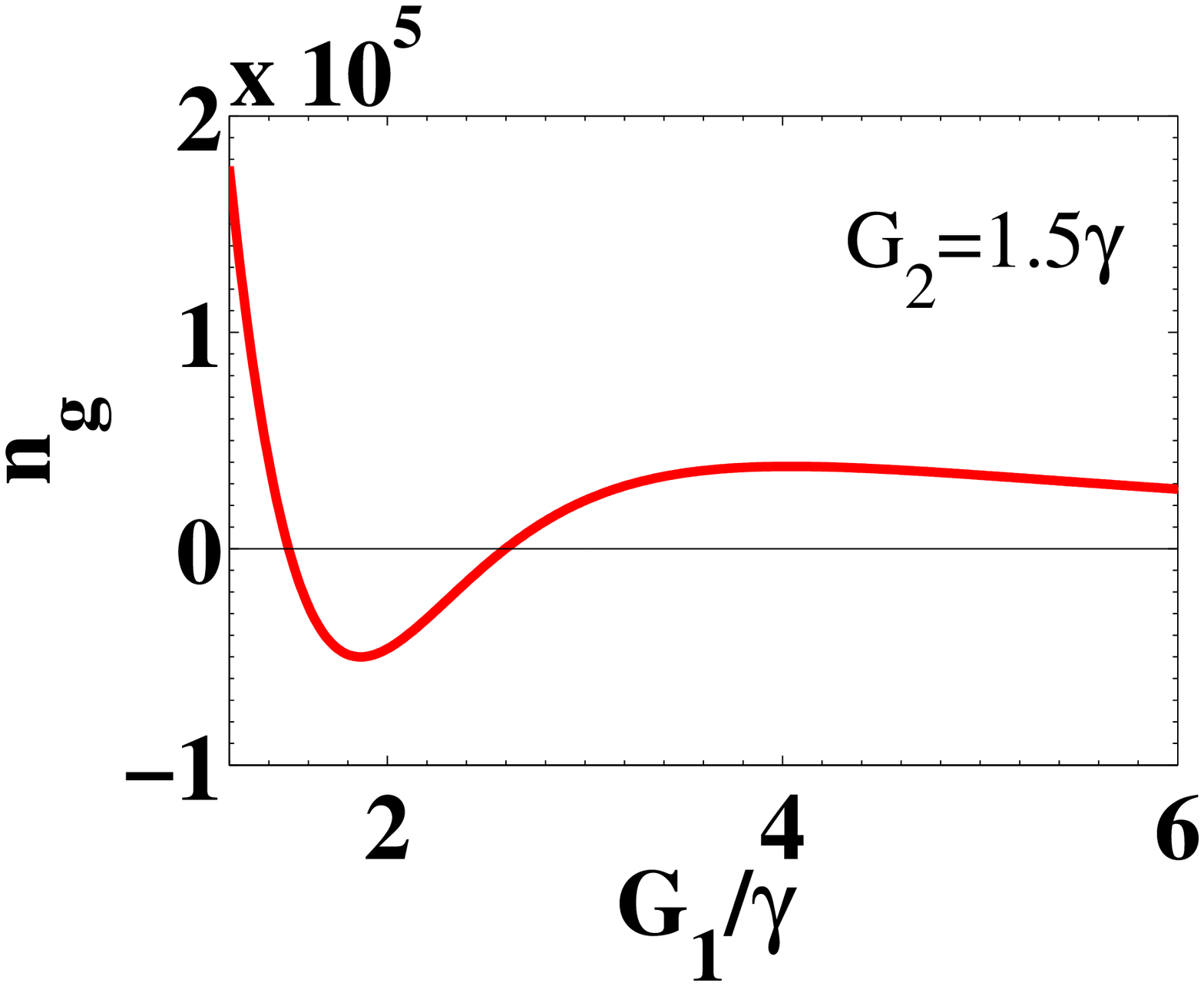}} & \subfigure[]{\includegraphics[scale=0.23]{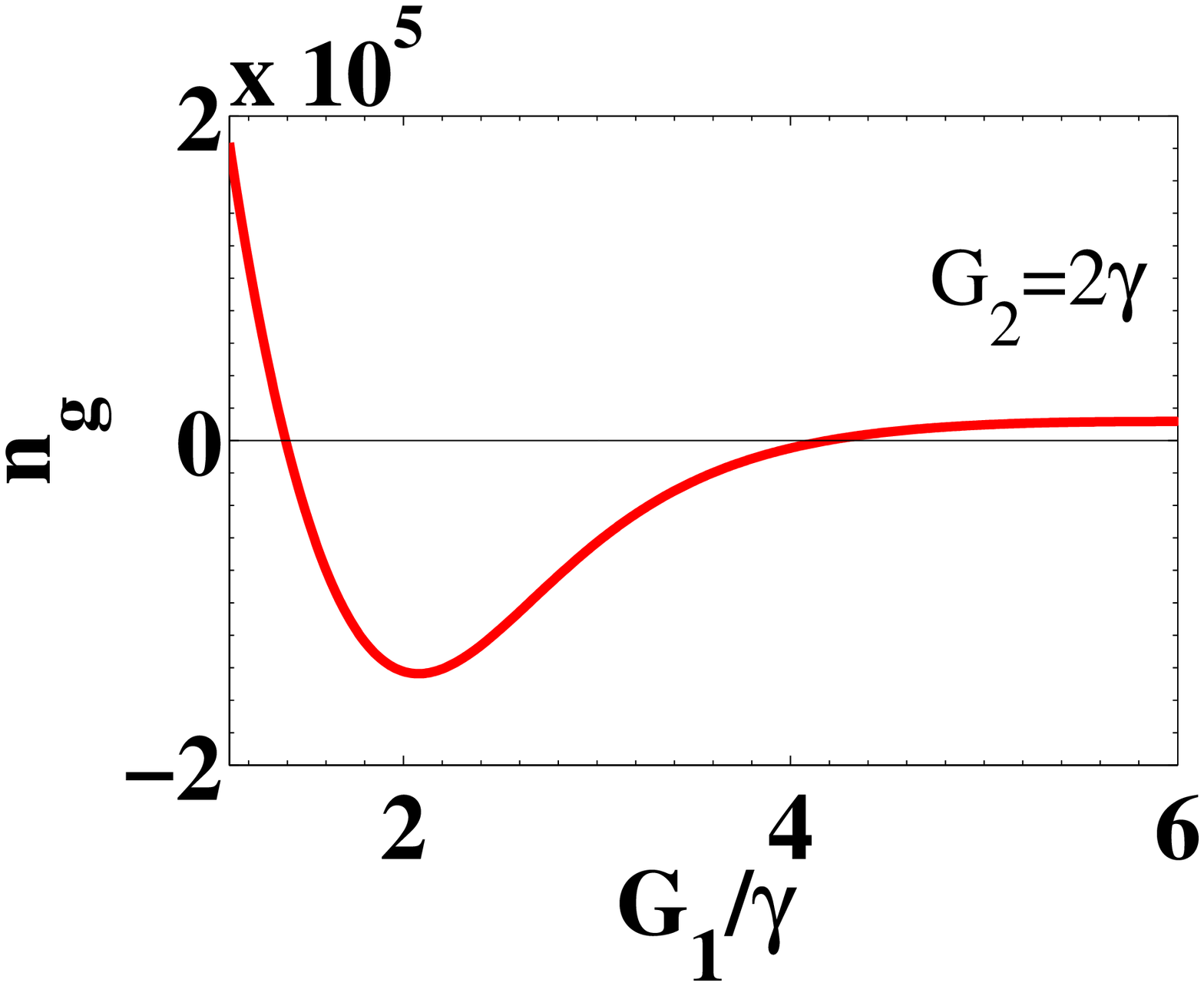}}\\
\end{tabular}
\end{center}
\caption{Variation of group index $n_g$ with Rabi frequency $G_1/\gamma$ for a particular value of $G_2/\gamma$. Other parameters are same as in  Figure 4.}
\label{fig(4)}
\end{figure}
\indent
When both the circular components of control field are switched on and are on resonance,  the coherences in (\ref{eq7}) are proportional to population differences as in (\ref{eq8}). The first order coherences are equal and are given by
\begin{equation}
\begin{array}{c}
\tilde{\rho}_{31}^{(-1)}=\tilde{\rho}_{42}^{(-1)}=\frac{\left[-irsu+\frac{\omega_{pc}\vert G\vert^2}{-i\Delta+\Gamma_{1}}(-\Gamma_{2}-\Gamma_{3}+2i\omega_{pc})\right]\tilde{\rho}_{pd}^{(0)}} {qrsu +2\vert G\vert^2 p(-\Gamma_{2}-\Gamma_{3}+2i\omega_{pc})}\\ \;
\end{array}
\label{eq9}
\end{equation}

where,
\begin{eqnarray}
\begin{array}{lll}
p &=& -\Gamma_{1}+i\omega_{pc}\\
q &=&   p-i\Delta\\
r &=& -\Gamma_{2}+i\omega_{pc}\\
s &=&  p+i\Delta\\
u &=& -\Gamma_{3}+i\omega_{pc}\\ \;
\end{array}
\label{eq10}
\end{eqnarray}
Both $\tilde{\rho}_{31}^{(-1)}$ and $\tilde{\rho}_{42}^{(-1)}$ exhibit similar spectrum. The variation of real and and imaginary parts of (\ref{eq4}), as a combined effect of these coherence terms, with respect of the probe detuning $\delta/\gamma$ (Note that the parameter $\gamma$ is defined as $\gamma=\frac{A}{6}$, taking all the decays from $\vert 3\rangle, \vert 4\rangle$ to $\vert 1\rangle, \vert 2\rangle$ and $A=2\pi \times 6.079$ MHz is the Einstein's A-coefficient in $^{39}$K.) are shown in Figure 3.
For resonant control field, i.e. when $\Delta=0$, the dispersion profile becomes anti-symmetric and the absorption profile becomes symmetric around $\delta=0$.  When $G=0.5\gamma$ (that corresponds to a power of 0.415 mW/cm$^{2}$, in case of D$_1$ transition in $^{39}$K and $\gamma=2\pi \times1.0132$ MHz), the absorption is zero and dispersion has a positive slope [Figure 3(a)]. The absorption spectrum continues to  exhibit two peaks as in the case of EIT until  $G=\gamma$ (corresponding to a power of 1.66 mW/cm$^2$). The separation between these two peaks increases with $G$  and hence the slope of dispersion profile decreases as shown in Figure 3(b). But at $G=1.5\gamma$ (3.37 mW/cm$^2$), the slope of dispersion becomes zero at zero probe detuning, as shown in Figure 3(c). With further increase in the control field strength to $G=2\gamma$ (6.64 mW/cm$^2$), the slope of the dispersion profile becomes negative, while the absorption remains zero near probe field resonance, as shown in Figure 3(d). Note that the absorption profile for $G\geq 1.5\gamma$ is similar to Mollow spectrum in a driven two-level system but the dispersion profile behaves in a distinct way, compared to such a system. We have found that for larger $G$, this slope remains negative, along with zero absorption \cite{wang2006}. The physical explanation of the zero absorption and gain peaks has been provided in Section 4.  We emphasize that the change in slope from positive to negative is a combined effect of both the circular components of the control field. If one of the circular component of the control field could be switched off, the slope would remain positive as in EIT.

As the slope of dispersion profile changes from normal to anomalous, there is a corresponding switching of group velocity of the probe field from the subluminal domain to the superluminal, at the resonance. The group index $n_g$ is related to the group velocity $v_g$ through the following relation:
\begin{equation}
v_g = \frac{c}{n_g}= c\left[1+2\pi \mbox{Re} \chi_{\pi}(\omega_{p})+2\pi\omega_{p}\frac{\partial}{\partial{\omega_{p}}} \mbox{Re} \chi_{\pi} (\omega_{p})\right]^{-1}\;,
\label{eq11}
\end{equation}
where $c$ is the velocity of light in vacuum and $\omega_p$ is the frequency of the probe field. Thus by modifying the group index, group velocity can be made subluminal ($n_g\gg 1$, or $v_g\ll c$), superluminal ($n_g < 1$, or $v_g > c$) or even negative ($n_g<0$, or $v_g<0$). When Re$\chi_{\pi}$ is negligible, then the slope of the dispersion profile at resonance plays a major role in the determination of group index and group velocity. Note that the real part of $\chi_\pi$ is zero at probe resonance, as shown in Figure 3. In Figure 4, we show the variation of $n_g$ with $G/\gamma$ for $\Delta=0$. Clearly, the group index changes from a large positive value to a large  negative value as the strength of the control field is varied. It is positive for $G<1.5\gamma$, zero at $G=1.5\gamma$, and becomes negative for $G>1.5\gamma$ .  The variation of group index for $G<1.5\gamma$ is shown in the insets (a) and (b) in Figure 4. The group index of the order of $10^7$ for coupling power of $12$mW/cm$^{2}$ is reported in \cite{harris1999}. But in our model, for smaller value of $G$($\approx$ 0.1$\gamma$ $\approx$0.017 mW/cm$^{2}$), group index can be as large as 3.2 $\times 10^{8}$. Further, the $n_g$ attains the largest negative value $-1.5\times10^{5}$ at $G=2.4\gamma$ (9.56 mW/cm$^2$). 

\subsection{Case when $G_1\ne G_2$}

We further consider the situation when $G_1\neq G_2$ that can be achieved by using a variable neutral density filter (see Figure 2). We have solved the density matrix equations analytically (see Appendix A for the solutions) for $\rho_\pi$ in this case. We show in Figure 5 the variation of $n_g$ with $G_1/\gamma$ for a particular value of $G_2/\gamma$. It is clear that for $G_2<1.5\gamma$, $n_g$ remains positive for all values of $G_1/\gamma$. But for $G_2\geq 1.5\gamma$, $n_g$ first switches from positive values to the negative and then from negative values to the positive, as $G_1/\gamma$ is further increased. This suggests that for $G_2\geq 1.5\gamma$, the group velocity of the probe field switches from subluminal to superluminal and again to subluminal, as Rabi frequency $G_1$ of the control field component is increased.  We thus obtain a ``valley of anomaly", in which the dispersion remains anomalous and the group index remains negative. This demonstrates a {\it better coherent control} over group velocity, all for small values of the strength of the control field. In Figure 6, we show the variation of $n_g$ with $G_1/\gamma$ and $G_2/\gamma$. The region of largest superluminality appears around $G_1=G_2=G\approx 2.3\gamma$. We also obtain the largest subluminality at $0<G<\gamma$ (region not shown).
In addition, we can have two sets of values of control field components, for example, $G_1=G_2=1.5\gamma$ and $G_1=2.6\gamma, G_2=1.5\gamma$, at which the absorption in the medium is zero and $n_g=1$. In these cases, the system clearly appears as vacuum to the probe pulses.

\begin{figure}[ht!]
\centering
\includegraphics[scale=0.4]{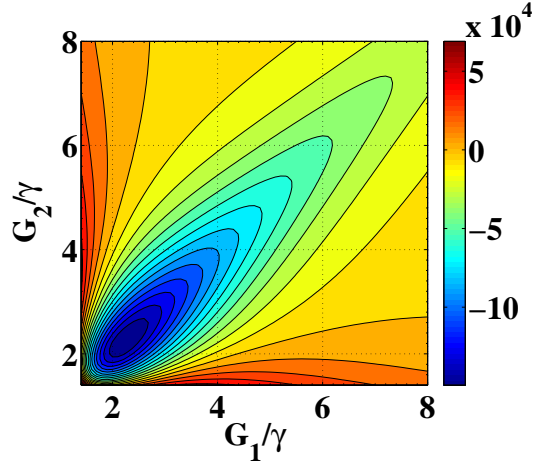}
\caption{Contour plot for $n_g$ with respect to $G_1/\gamma$ and $G_2/\gamma$ with the parameters same as in Figure 4.}
\label{fig(5)}
\end{figure}

\subsection{Dynamics of pulse propagation}

To confirm these results, we next study the propagation of a Gaussian pulse with a normalized envelope given by
\begin{eqnarray}
\begin{array}{cc}
\varepsilon(\omega)=\varepsilon_{0} \frac{1}{\sigma \sqrt{\pi}} \exp{\left[-\omega^{2}/ \sigma^{2} \right]}; & \varepsilon(t)=\varepsilon_{0}\exp{\left(-\sigma^{2}t^{2}/4\right)}
\end{array}
\label{eq12}
\end{eqnarray}
where, $\sigma$ gives the width of the pulse in frequency domain. The unwanted absorption (gain) of different frequency components of the pulse can be avoided by choosing a narrow-band pulse, so that it remains well within the EIT window (gain doublet). For numerical calculation, we choose $\sigma = 2\pi \times 5$ kHz. In Figure 7, we show the numerical results of the pulse (\ref{eq12}) at the output of the medium for several values of $G$. The pulse delay of 30 $\mu s$ for $G$ = $\gamma$ is shown in Figure 7(a). The group index calculated from the delay between the reference pulse and the output pulse is in good agreement, with the value of the group index ($n_g$ = 9.8$\times$10$^{5}$) as shown in Figure 4, for $G$ = $\gamma$.  Figure 7(b) shows that medium behaves like vacuum ($n_g$ = 1) for the input pulse, as the relative delay between the reference pulse and the output pulse vanishes as predicted in Figure 4 for $G$ = 1.5$\gamma$. The pulse advancement of -4.71 $\mu s$ at $G$ = 2$\gamma$ is demonstrated in Figure 7(c) and it corresponds to a group index of -1.4$\times$10$^{5}$, which also agrees well with the Figure 4 at $G$ = 2$\gamma$.

 \begin{figure}[!htb]
\begin{center}
\subfigure[]{\includegraphics[scale=0.23]{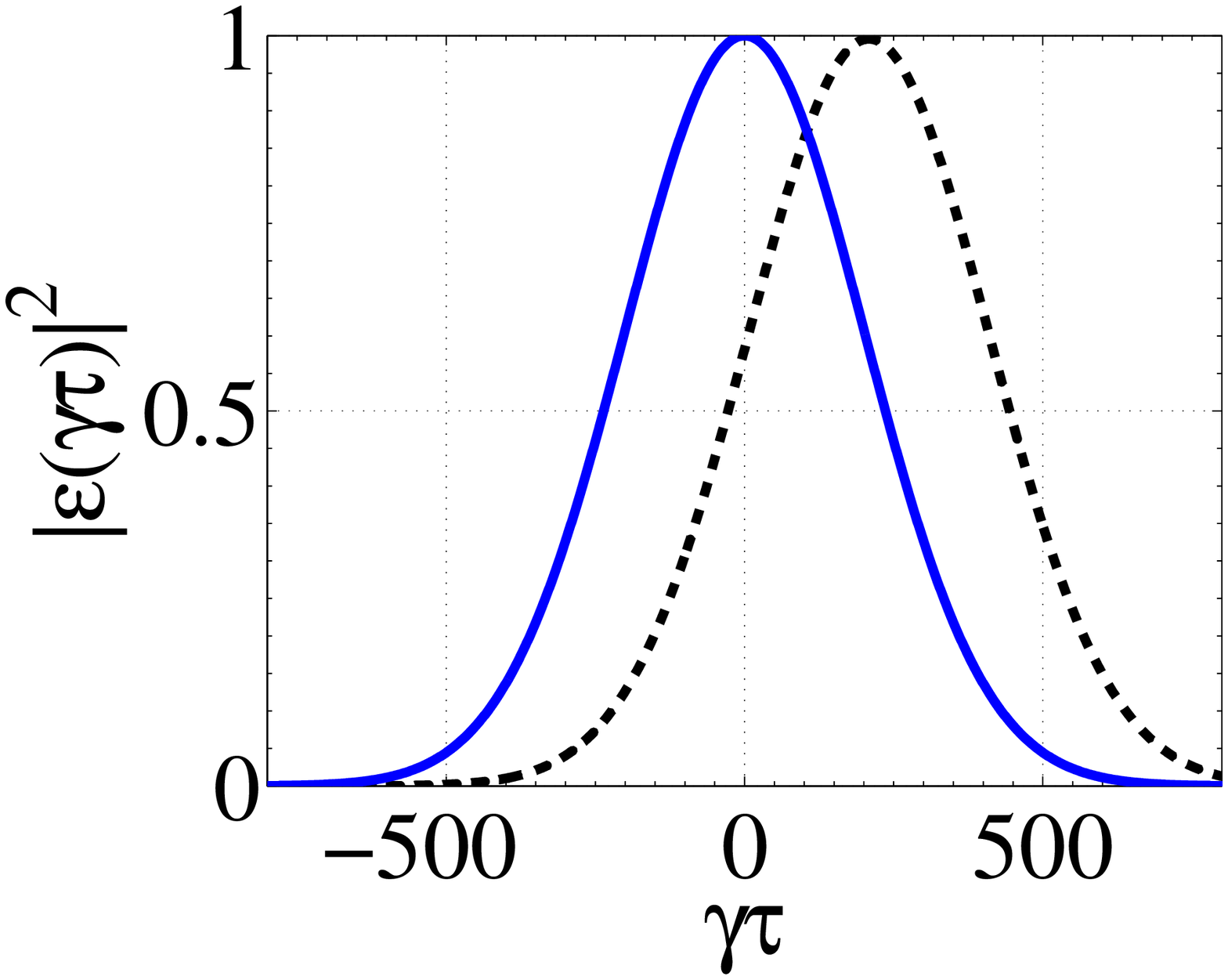}} \hspace{-1mm}  \subfigure[]{\includegraphics[scale=0.23]{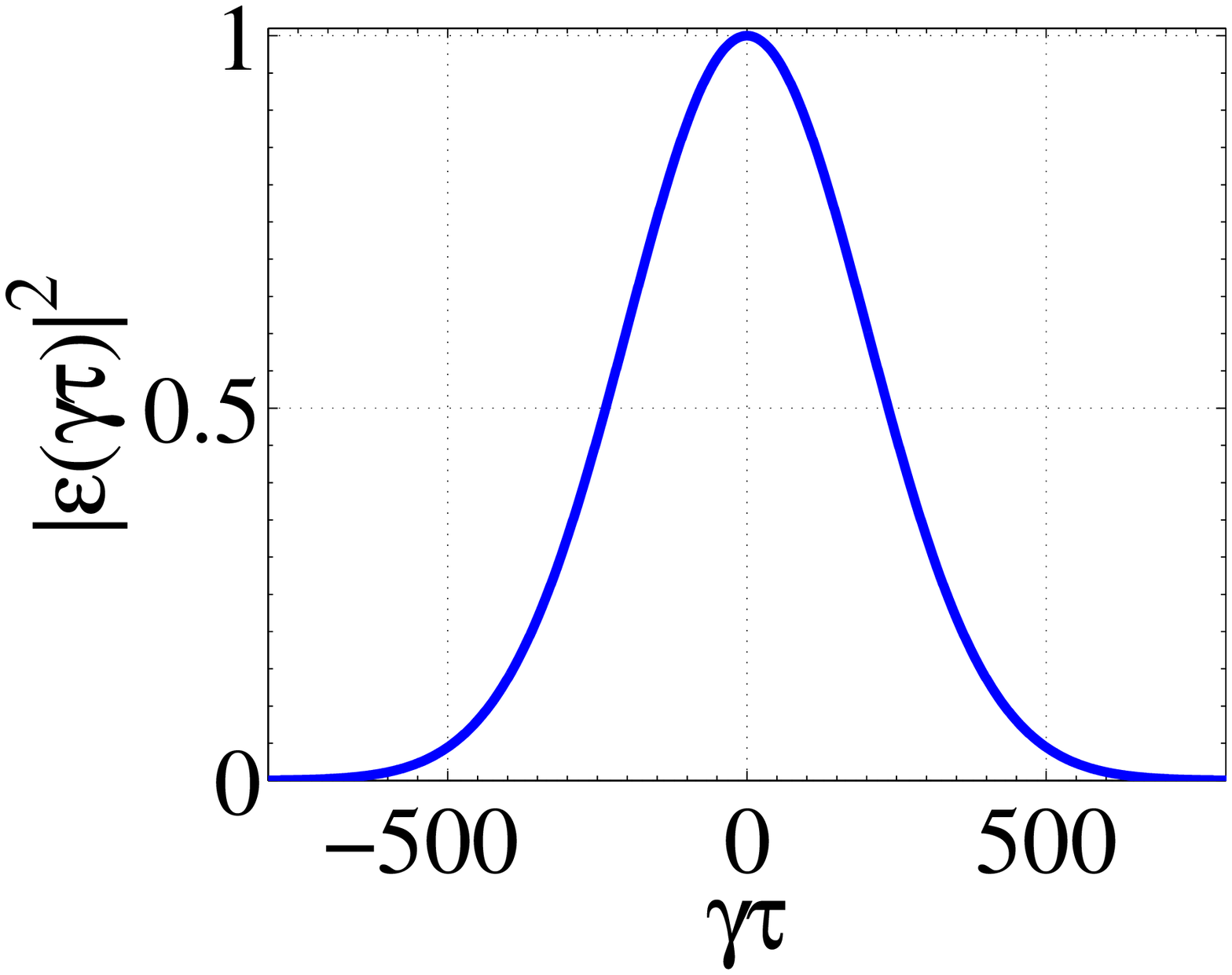}}\\
 \subfigure[]{\includegraphics[scale=0.23]{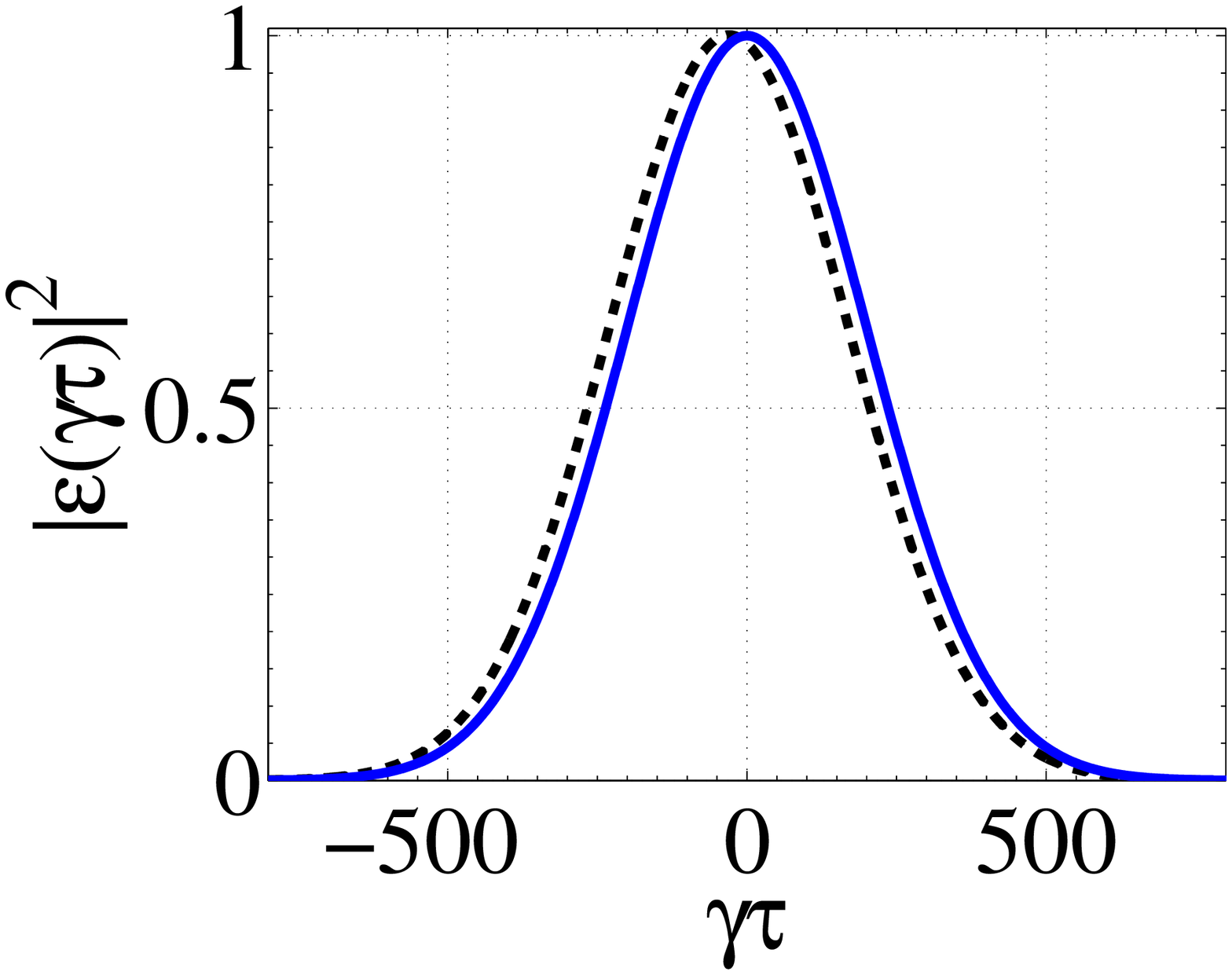}}
\end{center}
\caption{Output pulse after traveling by $L=1$ cm through vacuum (solid line) and the atomic medium (dotted line) of same length for  (a) $G$ = $\gamma$, (b) $G$ = 1.5$\gamma$, and (c) $G$ = 2$\gamma$. The pulse width $\sigma$ is chosen to be 2$\pi$ $\times$ 5 kHz. Other parameters are same as in Figure 4 and $\tau = t-L/c$.}
\label{fig(6)}
\end{figure}
\section{Discussions}

In a driven two level system \cite{ficek},  the probe field propagates through an anomalously dispersive medium without any absorption at probe resonance. This can be understood in terms of the states, partially dressed by the strong field. The transition dipole moment matrix elements among these dressed states cancel each other, to nullify the absorption of the probe field.  Similarly, in a three level system with $\Lambda$ configuration, a strong control field makes the system transparent to a weak probe field, thanks to EIT, which is associated to normal dispersion. The total transition amplitude from the ground state to the partially dressed states vanishes, leading to the zero absorption for the probe field. In this paper, we deal with a driven four-level system, where the absorption remains zero at probe resonance. This also can be understood in terms of the partial dressed states, arising due to the resonant control fields, which are given by
\begin{eqnarray}
\vert \pm\rangle_{14}&=&\frac{1}{\sqrt{2}}(\vert 1\rangle \pm\vert 4\rangle) \;,\nonumber\\
\vert \pm\rangle_{23}&=&\frac{1}{\sqrt{2}}(\vert 2\rangle \pm\vert 3\rangle) \;.
\label{dressed}
\end{eqnarray}
with  the respective eigen energies $\omega_{14}^{\pm} = \omega_{23}^{\pm} =\pm \vert G \vert$.
The absorption probability of the probe fields in the transitions $|1\rangle\leftrightarrow |3\rangle$ and $|2\rangle\leftrightarrow |4\rangle$ is proportional to the square of the total  transition amplitude.
This transition amplitude will be the sum of those in the transitions $\vert \pm\rangle_{23}\leftrightarrow\vert \pm\rangle_{14}$ as given by
\begin{equation}
\sum_{\alpha,\beta  \in \pm} {}_{23}\langle \alpha\vert\hat{\mu}\vert \beta \rangle_{14}\approx 2\mu_{21} \;,
\end{equation}
where  $\hat{\mu}$ is the dipole moment operator.
As the transition between level $\vert 1\rangle$ to $\vert 2\rangle$ is dipole forbidden, $\mu_{21}$ is zero. Thus the total transition amplitude and hence the absorption vanishes at probe resonance.

To understand the physics behind the tunability of group velocity, let us consider the absorption of the probe field in the $|1\rangle\rightarrow|3\rangle$ transition. When the $G_2$ field is on, the absorption profile gets split into two peaks, corresponding to the transitions $|1\rangle\rightarrow |\pm\rangle_{23}$. In this case, at resonance, the absorption probability of the probe photon gets nullified with the gain probability from the dressed states $|\pm\rangle_{23}$ (note that the state $|2\rangle$ is also initially populated). We obtain a classic case of EIT, leading to normal dispersion at resonance. If the $G_1$ field is now switched on, some of the population from the state $|1\rangle$ is pumped to $|2\rangle$ via $|4\rangle$, and it thus increases the gain probability. If $G_1$ is increased beyond certain threshold value, the gain of the probe field dominates over the absorption in the $|1\rangle\leftrightarrow |3\rangle$ transition, at near-resonance domain. The gain is nullified at resonance, that generates two gain peaks. We thus obtain a gain doublet that is associated with anomalous dispersion \cite{wang2000}. Note that the total susceptibility of the medium is a sum over those in the two probe transitions (\ref{eq4}). Therefore, for much larger values of $G_1$ (keeping $G_2$ constant), the system starts to behave like an EIT system, with reference to the probe field in the $|2\rangle\leftrightarrow |4\rangle$ transition. This again provides two absorption peaks, associated with normal dispersion at resonance. In this way, by adjusting the amplitude of the two polarization components of the control field, one can  switch the dispersion from normal to anomalous and then back to normal. It may be noted that for $G_1\gg G_2$, the absorption peaks becomes far apart (separation between them is of the order of $|2G_1|$), and the slope of the susceptibility also tends to vanish. This explains why $n_g$ tends to unity for large $G_1$, as shown in Figure 5.\\

\begin{figure}[!htb]
\begin{center}
\subfigure[]{\includegraphics[scale=0.23]{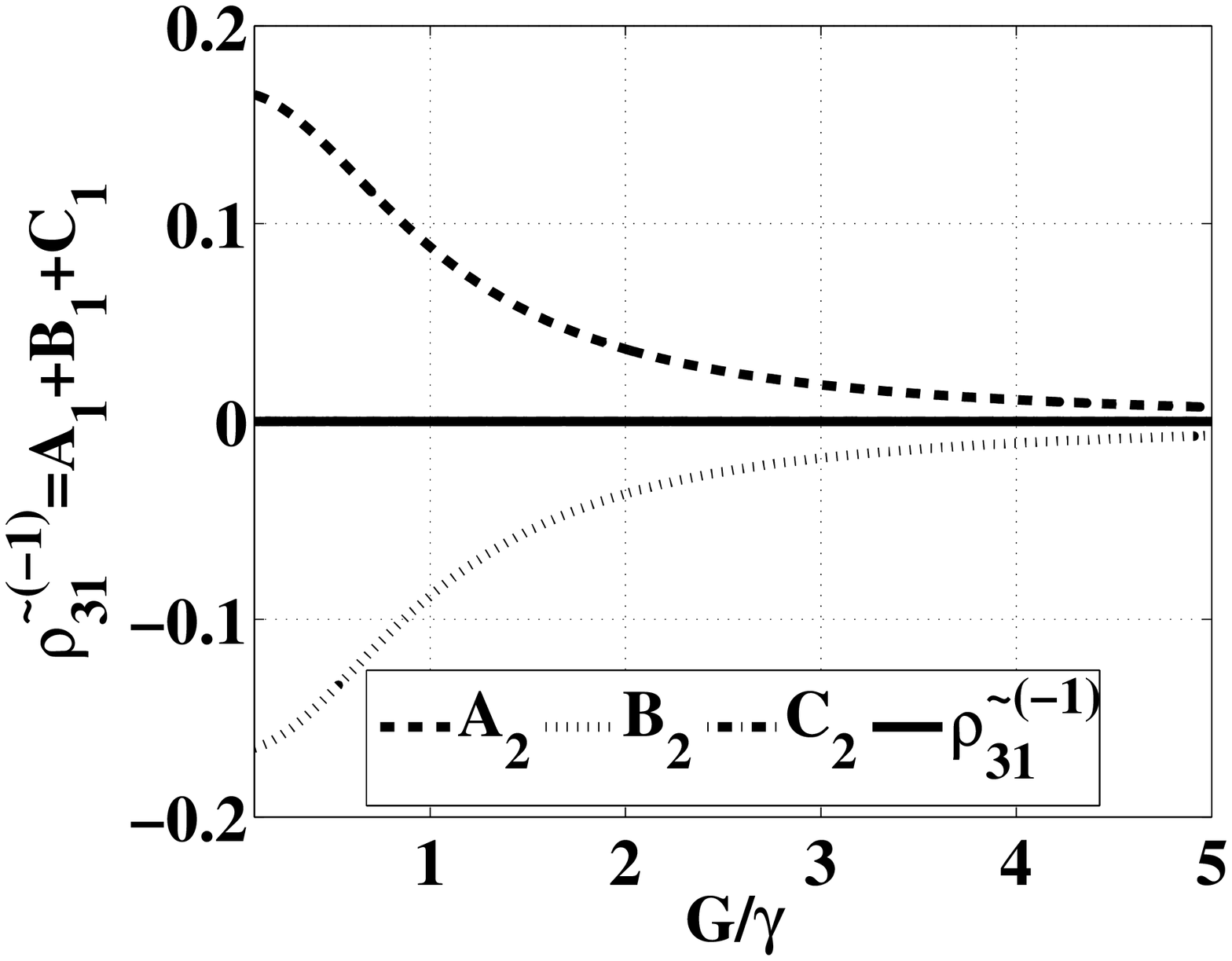}} \hspace{-1mm}  \subfigure[]{\includegraphics[scale=0.23]{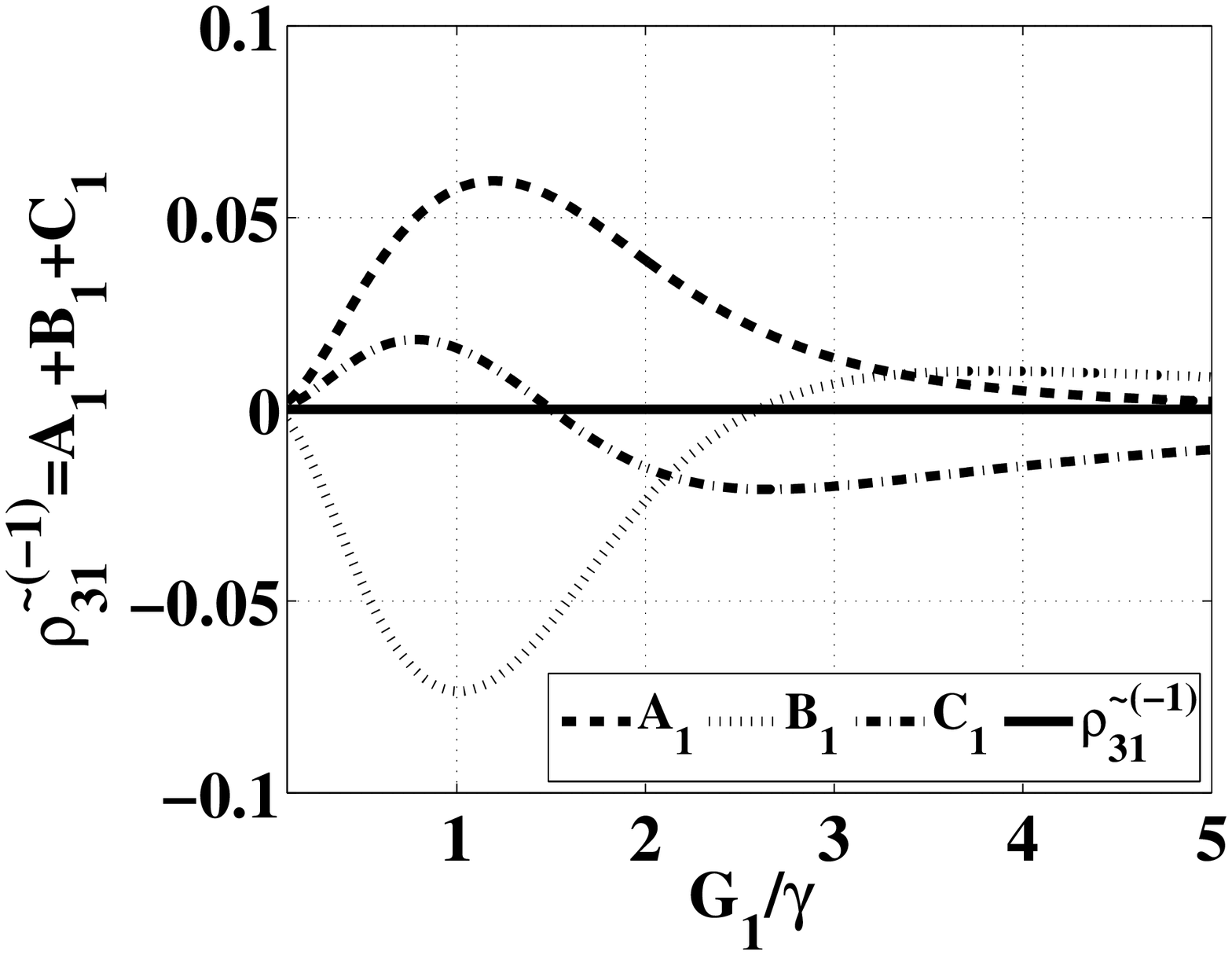}}\\
 \subfigure[]{\includegraphics[scale=0.23]{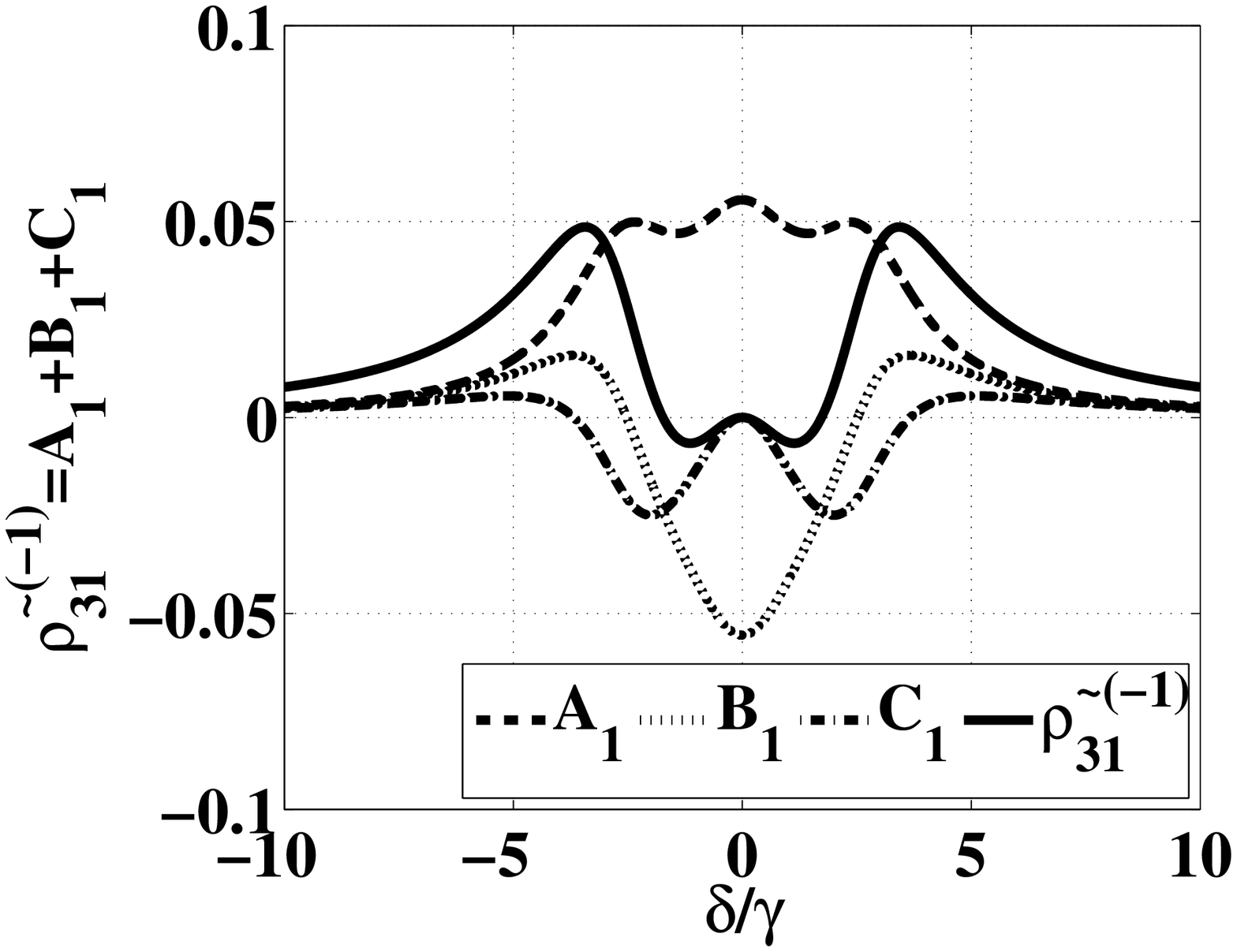}}
\end{center}
\caption{Variation of imaginary parts of the terms  $A_1$ (thick dashed line), $B_1$ (thin dashed line),  $C_1$ (dot-dashed line), and $\tilde{\rho}_{31}^{(-1)}$ (solid line) with (a) $G/\gamma$ (b) $G_1/\gamma$ for $G_2=1.5\gamma$ (c) $\delta/\gamma$ for $G=1.5\gamma$, where $A_1$, $B_1$, and $C_1$ are given by (\ref{eqb3}) in Appendix A. Other parameters are the same as in  Figure 4.}
\label{fig(6)}
\end{figure}
Further, as shown in the Appendix A, the probe coherence $\tilde{\rho}_{31}^{(-1)}$ is a sum of three zeroth-order contributions: population-difference term $A_1$ (proportional to $\tilde{\rho}_{11}^{(0)}-\tilde{\rho}_{33}^{(0)}$) and coherence terms $B_1$ (proportional to $\tilde{\rho}_{23}^{(0)}$) and $C_1$ (proportional to $\tilde{\rho}_{14}^{(0)}$).  In Figure 8, we show the individual contributions of these terms in the absorption spectra. At two photon resonance ($\delta=\Delta$), the contributions from two terms  $A_1$ and $B_1$ to the absorption spectra for $\tilde{\rho}_{31}^{(-1)}$ cancel each other, while the other coherence term
$C_1$ remains zero for $G_1=G_2=G$ [Figure 8(a)], leading to no absorption.  When $G_1\ne G_2=1.5\gamma$,  the transparency at two photon resonance is caused by all the coherence terms $B_1$ and $C_1$, as well as the term $A_1$ [Figure 8(b)].

The gain in certain frequency domain can be further understood in terms of competing effect of the coherence terms and population difference terms. As shown in Figure 8(c), the negative contributions from $B_1$ and $C_1$ are larger in magnitude than the positive contribution from $A_1$ at around $\delta=\pm G$. Thus we obtain a situation where a gain-doublet and therefore a window of anomalous dispersion is generated by control-field-induced coherence.

\section{Effect of Doppler broadening on group index}
The Doppler broadening for the copropagating control and probe fields can be accounted by replacing $\omega_{c}$ by ($\omega_{c}+kv$) and $\omega_{p}$ by ($\omega_{p}+kv$), where, $\vec{v}$ is the velocity of an atom.  As the pump and probe field are copropagating, the two-photon resonance condition remains unaffected by the Doppler broadening, as the velocity-dependent detunings $\Delta_{v}=\Delta+kv$ and $\delta_{v}=\delta+kv$ maintains the relation $\Delta_{v}-\delta_{v}=\Delta-\delta$.
The susceptibility (\ref{eq5}) is averaged over the Maxwell-Boltzmann distribution of velocities at a temperature $T$
\begin{equation}
\sigma_{D}(kv)=\frac{1}{\sqrt{2\pi\omega_{D}^{2}}}\exp{\left[-(kv)^{2}/2\omega_{D}^{2}\right]},
\end{equation}
Here $\omega_{D}$ is the Doppler width defined by
\begin{equation}
\omega_{D}=\sqrt{K_{B}T\omega_{p}^{2}/Mc^{2}}
\end{equation}
$K_B$ is the Boltzmann constant and M is the mass of an atom.
The Doppler-averaged susceptibility can be written as
\begin{equation}
\langle \chi(kv)\rangle = \int_{-\infty}^{+\infty} \chi(kv)\sigma_{D}(kv)\,d(kv)
\end{equation}
The group index in such a medium takes the following form:
\begin{equation}
\langle n_g \rangle=1+2\pi \mbox{Re}\left[\langle \chi_{\pi}(\omega_p)\rangle\right]+2\pi \omega_{p} \frac{\partial}{\partial \omega_p}\mbox{Re}\left[\langle \chi_{\pi}(\omega_p)\rangle\right].
\end{equation}

\begin{figure}[!htb]
\begin{center}
\subfigure[]{\includegraphics[scale=0.2]{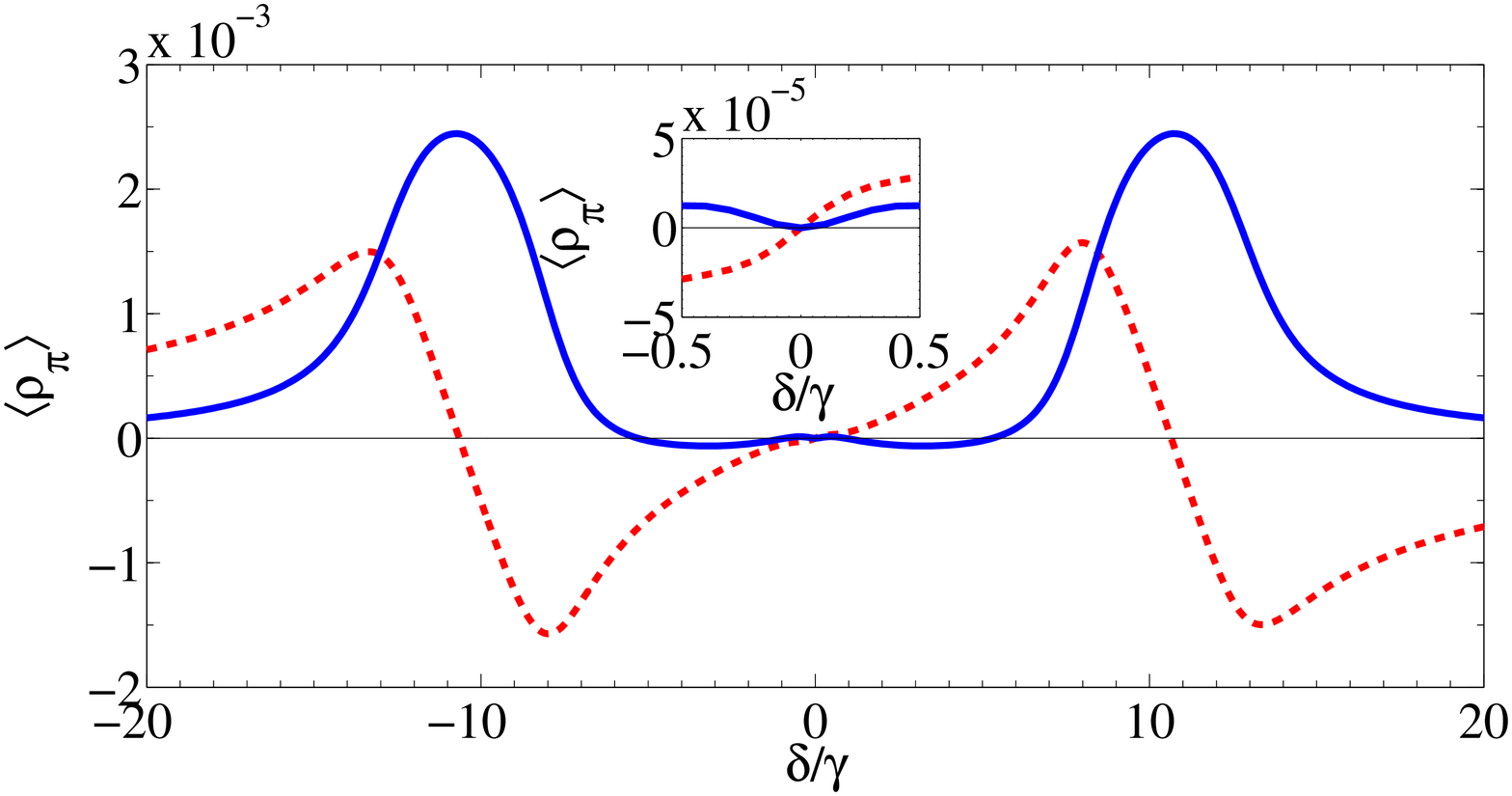}} \hspace{-10mm}  \subfigure[]{\includegraphics[scale=0.2]{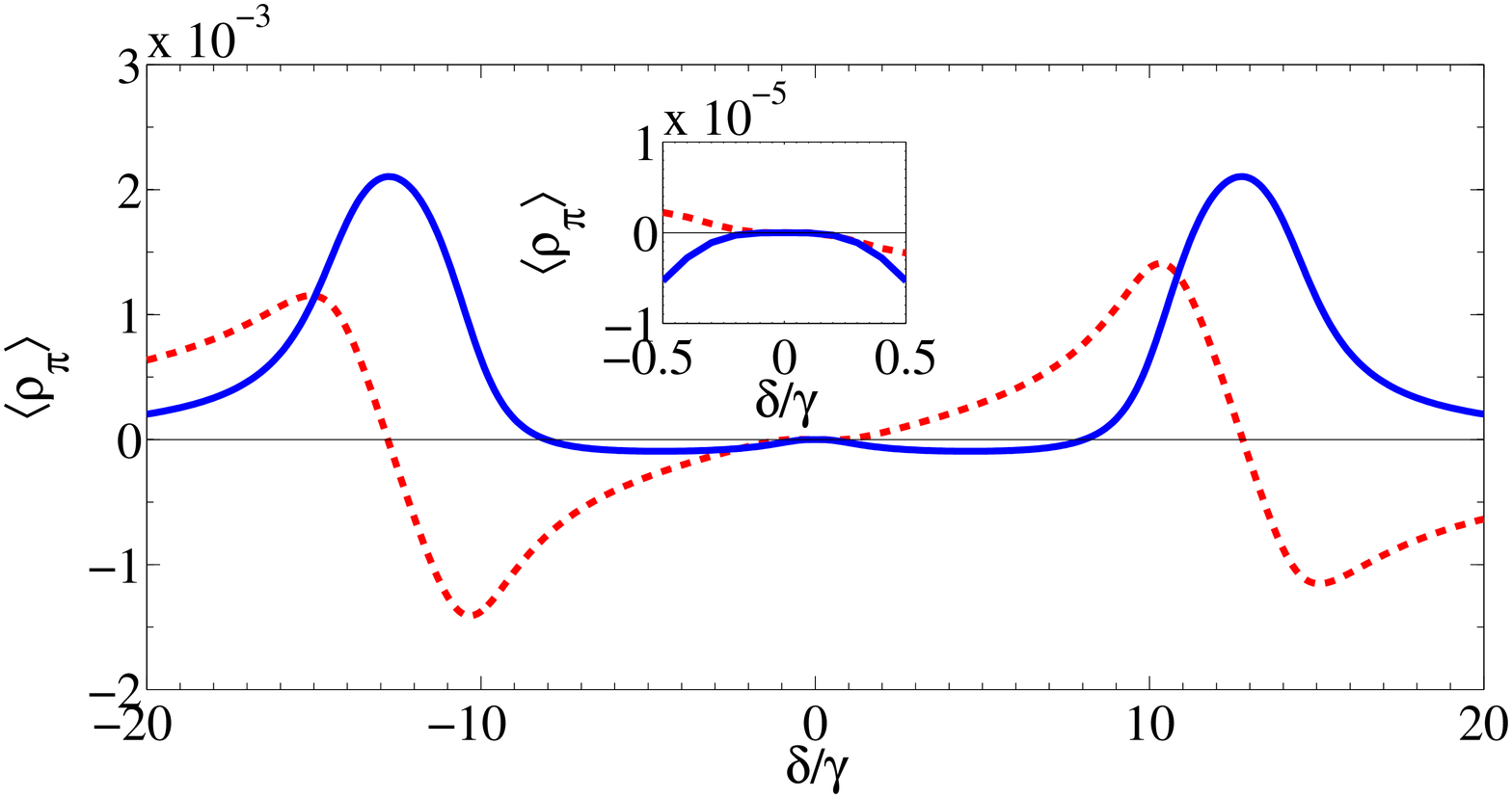}}\\
 \subfigure[]{\includegraphics[scale=0.2]{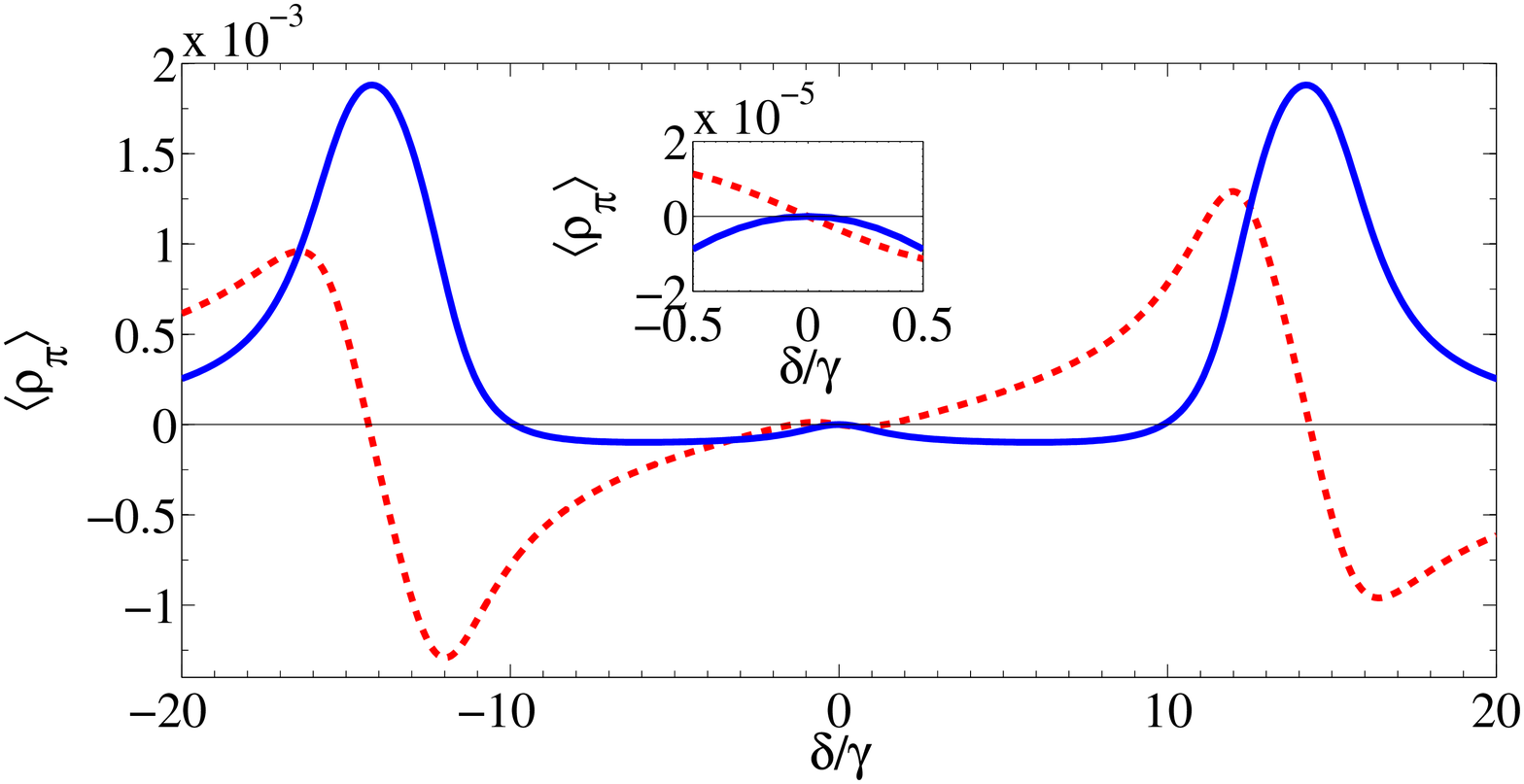}}
\end{center}
\caption{The variation of real (dotted line) and imaginary part (solid line) of $\langle\rho_{\pi}\rangle$ with the detuning of the probe field $\delta/\gamma$ with (a) $G$ = 4$\gamma$, (b) $G$ = 5.18$\gamma$ and (c) $G$ = 6$\gamma$. The insets show the magnified part of the same. The Doppler width is chosen as $\omega_D$ = 324$\gamma$ at room temperature and other parameters are same as in Figure 4.}
\label{fig(8)}
\end{figure}
We show in Figure 9, the numerical results for the real and imaginary parts of $\langle\rho_{\pi}\rangle$ for resonant control field. We choose Doppler width $\omega_{D}$ = 324$\gamma$ at room temperature. The imaginary part of $\langle\rho_{\pi}\rangle$ still remain zero at probe resonance, whereas the slope of the real part changes from normal to anomalous by increasing the strength of the control field.  In Figure 10, we show the numerical results for the behavior of $\langle n_{g}\rangle$ as a function of $G$, for resonant probe field. In Doppler broadened medium, the switching from subluminal to superluminal group velocity is still seen to occur but at $G$ = 5.18$\gamma$.

\begin{figure}[ht!]
\centering
\includegraphics[scale=0.4]{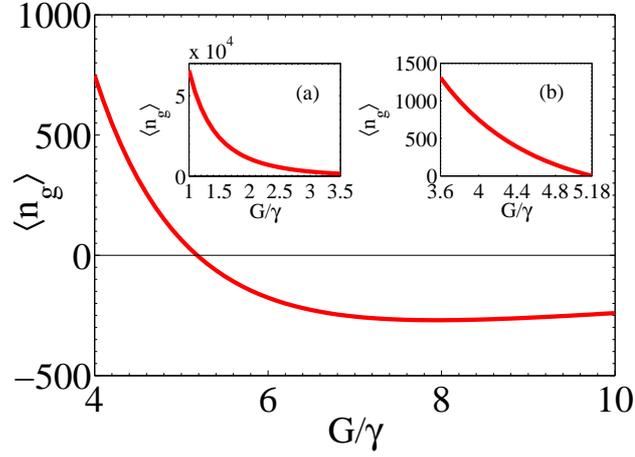}
\caption{Variation of group index $\langle n_{g}\rangle$ in Doppler broadened medium with $G/\gamma$. Insets (a) and (b) shows the variation of $\langle n_{g}\rangle$ for 1$<G/\gamma<$3.5 and 3.6$<G/\gamma<$5.18, respectively. Other parameters are same as in Figure 9.}
\label{fig(5)}
\end{figure}
\section{Conclusion}
In conclusion, we have discussed the possibility of tunable group velocity in $J=\frac{1}{2}\leftrightarrow J=\frac{1}{2}$ transition in alkali atoms, driven by a single linearly polarized resonant control field. We find that the group velocity of a linearly polarized probe field can be switched from positive (subluminal propagation) to negative (superluminal propagation) by changing the strength of the control field. By suitably manipulating the values of $G_1$ ($G_2$),  one also can obtain a range of $G_2$ ($G_1$), where the superluminality persists. We refer to this range as a ``valley of anomaly", in which the probe absorption remains zero. We demonstrate, using detailed analytical and numerical results, that by keeping the power of the circular components of control field  at low values, one can scan over a large range of group index (from $\sim +10^{8}$ to $\sim -10^{5}$). We provide detailed physical explanation of the gain and transparency in terms of coherence, created by the control field components. We further show how a Gaussian pulse propagates through such a medium and attains different group velocities at different control fields. We provide detailed numerical analysis for the group index variation by considering the Doppler broadening at room temperature. We also have proposed a possible experimental set-up to demonstrate our results.

It must be noted that the effect of hyperfine structure in alkali atoms cannot be neglected. For example, in $^{39}$K the hyperfine splitting of the $2P_{1/2}$ state is 55.5 MHz and of the $2S_{1/2}$ state is 461.7 MHz, leading to non-degenerate excited and ground state manifolds. One could use the $|F=1,m_F=+1\rangle$ and $|F=2,m_F=+2\rangle$ as relevant levels in both the manifolds. Such highest order magnetic hyperfine sublevels can be populated by using optical pumping, as routinely done in cold atom experiments. However, if the probe field is kept near-resonant to one of the $\Delta m_F=0$ transition, it will remain highly off-resonant in the other $\Delta m_F=0$ transition. Therefore, it may not be possible to obtain the valley of anomaly (as described in Section 4) by increasing the control field strength. 

\appendix
\section{General first order solutions of the density matrix equations (\ref{eq2})}
We here provide the solutions of the equations (\ref{eq2}), up to first order in $g$, when the  control fields have different amplitudes, i.e., when $G_1\ne G_2$. The coherence terms can be written as
\begin{eqnarray}
\tilde{\rho}_{31}^{(-1)} &=& A_1+B_1+C_1\\
\tilde{\rho}_{42}^{(-1)} &=& A_2+B_2+C_2
\end{eqnarray}
where,

\begin{equation}
\begin{array}{lll}
A_1 &=& -i\frac{[rsu+\vert G_2\vert^{2}r+\vert G_1\vert^{2}u](\tilde{\rho}_{11}^{(0)}-\tilde{\rho}_{33}^{(0)})}{qrsu+\vert G_2\vert^{2}(qr+su) +\vert G_1\vert^{2}(qu+rs)+(\vert G_1\vert^{2}-\vert G_2\vert^{2})^2}\\
B_1 &=& \frac{(su+\vert G_2\vert^{2}-\vert G_{1}\vert^{2})G_{2}\tilde{\rho}_{23}^{(0)}}{qrsu+\vert G_2\vert^{2}(qr+su) +\vert G_1\vert^{2}(qu+rs)+(\vert G_1\vert^{2}-\vert G_2\vert^{2})^2}\\
C_1 &=& \frac{(rs-\vert G_{2}\vert^{2}+\vert G_{1}\vert^{2})G_{1}\tilde{\rho}_{14}^{(0)}}{qrsu+\vert G_2\vert^{2}(qr+su) +\vert G_1\vert^{2}(qu+rs)+(\vert G_1\vert^{2}-\vert G_2\vert^{2})^2}\\
A_2 &=& -i\frac{[rsu+\vert G_1\vert^{2}r+\vert G_2\vert^{2}u](\tilde{\rho}_{22}^{(0)}-\tilde{\rho}_{44}^{(0)})}{qrsu+\vert G_2\vert^{2}(qu+rs) +\vert G_1\vert^{2}(qr+su)+(\vert G_1\vert^{2}-\vert G_2\vert^{2})^2}\\
B_2 &=& \frac{(su+\vert G_1\vert^{2}-\vert G_{2}\vert^{2})G_{1}\tilde{\rho}_{14}^{(0)}}{qrsu+\vert G_2\vert^{2}(qu+rs) +\vert G_1\vert^{2}(qr+su)+(\vert G_1\vert^{2}-\vert G_2\vert^{2})^2}\\
C_2 &=& \frac{(rs-\vert G_{1}\vert^{2}+\vert G_{2}\vert^{2})G_{2}\tilde{\rho}_{23}^{(0)}}{qrsu+\vert G_2\vert^{2}(qu+rs) +\vert G_1\vert^{2}(qr+su)+(\vert G_1\vert^{2}-\vert G_2\vert^{2})^2}
\end{array}
\label{eqb3}
\end{equation}

where $q,r,s,u$ are the same as in (\ref{eq10}).
\ack
The authors wish to acknowledge Dr. Rajesh V. Nair and Dr. K. C. Jena for useful discussions.


\section*{References}
\begin{thebibliography}{50}
\bibitem{harris1992} Harris S E, Field J E and  Kasapi A 1992 \textit{Phys. Rev. A} \textbf{46} R29.
\bibitem{harris1997}  Harris S E 1997 \textit{Phys. Today} \textbf{50(7)} 36.
\bibitem{harris1999} Hau L V, Harris S E, Dutton Z and Behroozi C H 1999 \textit{Nature (London)} \textbf{397} 594.
\bibitem{kash1999} Kash M M \textit{et al.} 1999 \textit{Phys. Rev. Lett.} \textbf{82} 5229 .
\bibitem{chien2001} Liu C, Dutton Z, Behroozi C H and Hau L V 2001 \textit{Nature (London)} \textbf{409} 490.
\bibitem{wang2000} Wang L J, Kuzmich A and Dogariu A 2000 \textit{Nature (London)} \textbf{406} 277.
\bibitem{kasapi1995} Kasapi A, Jain M, Yin G Y and Harris S E 1995 \textit{Phys. Rev. Lett.} \textbf{74} 2447.
\bibitem{schmidt1996} Schmidt O, Wynands R, Hussien Z and Meschede D 1996 \textit{Phys. Rev. A} \textbf{53} R27.
\bibitem{budker1999} Budker D, Kimball D F, Rochester S M and Yashchuk V V 1999 \textit{Phys. Rev. Lett.} \textbf{83} 1767 .
\bibitem{brillouin1960}  Brillouin L 1960 \textit{Wave propagation and Group Velocity} (Academic Press, New York).
\bibitem{born1997} Born M and Wolf E 1997 \textit{Principles of Optics} (Cambridge Uni. Press, Cambridge,7th ed.).
\bibitem{aggarwal2001} Agarwal G S, Dey T N and  Menon S 2001 \textit{Phys. Rev. A} \textbf{64} 053809.
\bibitem{arbiv2001} Bortman-Arbiv D, Wilson-Gordon A D and Friedmann H 2001 \textit{Phys. Rev. A} \textbf{63} 043818.
\bibitem{sahrai2004} Sahrai M, Tajalli H, Kapale K T and  Zubairy M S 2004 \textit{Phys. Rev. A} \textbf{70} 023813.
\bibitem{saaidi2011} Saaidi Kh, Ruzbahani B, Rabiei S W and Mahmoudi M 2011 \textit{Ar. J.  Phys.} \textbf{4(1)} 38.
\bibitem{Li2007} Li S, Yang X, Cao X, Xie C and Wang H 2007 \textit{J. Phys. B: At, Mol. Opt. Phys.} \textbf{40} 3211.
\bibitem{Zhang2007} Zhang Y, Brown A W and Xiao M 2007 \textit{Phys. Rev. Lett.} \textbf{99} 123603.
\bibitem{Yan2001} Yan M, Rickey E G and Zhu Y 2001 \textit{Phys. Rev. A} \textbf{64} 041801(R).
\bibitem{harris1998} Harris S E and Yamamoto Y 1998 \textit{Phys. Rev. Lett.} \textbf{81} 3611.
\bibitem{Kang2004}  Kang H, Hernandez G and Zhu Y 2004 \textit{Phys. Rev. A} \textbf{70} 011801.
\bibitem{Chen2009} Chen Y, Wei X G and Ham B S 2009 \textit{J. Phys. B: At, Mol. Opt. Phys.} \textbf{42} 065506.

\bibitem{Salloum2010}  Abi-Salloum T Y, Henry B, Davis J P and Narducci F 2010 \textit{Phys. Rev. A} \textbf{82} 013834.
\bibitem{Salloum2011}  Abi-Salloum T Y, Snell S, Davis J P and Narducci F A 2011 \textit{J. Mod. Opt.} \textbf{58} 2008.
\bibitem{s2006} Dasgupta S 2006 \textit{Opt. Commun.} \textbf{259} 683.
\bibitem{sankar2013} Davuluri S and Rostovtsev Y V 2013 \textit{J. Phys.} \textbf{414} 012005.
\bibitem{brown1997}  Brown W J,  Gardner J R,bGauthier D J and Vilaseca R 1997  \textit{Phys. Rev. A} \textbf{56} 3255.
\bibitem{santosh2013}  Kumar S, Laupr\^{e}tre T, Bretenaker F, Goldfarb F and Ghosh R 2013 \textit{Phys. Rev. A} \textbf{88} 023852.
\bibitem{lukin} Lukin M, Hemmer P R,  L\"offler M and Scully M O 1998 \textit{Phys. Rev. Lett.}  \textbf{81} 2675.
\bibitem{wu} Wu J H, Artoni M and La Rocca G C 2013 \textit{Phys. Rev. A}  \textbf{88} 043823.

\bibitem{hashmi2008} Hashmi F A and Bouchene M A 2008 \textit{Phys. Rev. Lett.} \textbf{101} 213601.
\bibitem{wang2006} This is unlike in Kai Wang,  Yang Gu and  Qi-Huang G 2007 \textit{Chinese Phys.} \textbf{16 (1)} 130, in which       the group index changes from positive to negative values and remains so with larger control fields.

\bibitem{ficek} Ficek Z and  Swain S 2005 \textit{Quantum Interference  and Coherence: theory and experiments} (Springer    series in optical sciences) p 213.
\end{thebibliography}
\end{document}